\PassOptionsToPackage{unicode}{hyperref}
\PassOptionsToPackage{hyphens}{url}
\PassOptionsToPackage{dvipsnames,svgnames,x11names}{xcolor}

\documentclass[
  12pt]{article}

\usepackage{amsthm}
\usepackage{amsmath,amssymb}
\usepackage{iftex}
\ifPDFTeX
  \usepackage[T1]{fontenc}
  \usepackage[utf8]{inputenc}
  \usepackage{textcomp} % provide euro and other symbols
\else % if luatex or xetex
  \usepackage{unicode-math}
  \defaultfontfeatures{Scale=MatchLowercase}
  \defaultfontfeatures[\rmfamily]{Ligatures=TeX,Scale=1}
\fi
\usepackage{lmodern}
\ifPDFTeX\else  
\fi
\IfFileExists{upquote.sty}{\usepackage{upquote}}{}
\IfFileExists{microtype.sty}{% use microtype if available
  \usepackage[]{microtype}
  \UseMicrotypeSet[protrusion]{basicmath} % disable protrusion for tt fonts
}{}
\makeatletter
\@ifundefined{KOMAClassName}{% if non-KOMA class
  \IfFileExists{parskip.sty}{%
    \usepackage{parskip}
  }{% else
    \setlength{\parindent}{0pt}
    \setlength{\parskip}{6pt plus 2pt minus 1pt}}
}{% if KOMA class
  \KOMAoptions{parskip=half}}
\makeatother
\usepackage{xcolor}
\makeatletter
\ifx\paragraph\undefined\else
  \let\oldparagraph\paragraph
  \renewcommand{\paragraph}{
    \@ifstar
      \xxxParagraphStar
      \xxxParagraphNoStar
  }
  \newcommand{\xxxParagraphStar}[1]{\oldparagraph*{#1}\mcal{}}
  \newcommand{\xxxParagraphNoStar}[1]{\oldparagraph{#1}\mcal{}}
\fi
\ifx\subparagraph\undefined\else
  \let\oldsubparagraph\subparagraph
  \renewcommand{\subparagraph}{
    \@ifstar
      \xxxSubParagraphStar
      \xxxSubParagraphNoStar
  }
  \newcommand{\xxxSubParagraphStar}[1]{\oldsubparagraph*{#1}\mcal{}}
  \newcommand{\xxxSubParagraphNoStar}[1]{\oldsubparagraph{#1}\mcal{}}
\fi
\makeatother

\usepackage{longtable,booktabs,array}
\usepackage{calc} % for calculating minipage widths
\usepackage{etoolbox}
\makeatletter
\patchcmd\longtable{\par}{\if@noskipsec\mcal{}\fi\par}{}{}
\makeatother
\IfFileExists{footnotehyper.sty}{\usepackage{footnotehyper}}{\usepackage{footnote}}
\makesavenoteenv{longtable}
\usepackage{graphicx}
\makeatletter
\def\maxwidth{\ifdim\Gin@nat@width>\linewidth\linewidth\else\Gin@nat@width\fi}
\def\maxheight{\ifdim\Gin@nat@height>\textheight\textheight\else\Gin@nat@height\fi}
\makeatother
\setkeys{Gin}{width=\maxwidth,height=\maxheight,keepaspectratio}
\makeatletter
\def\fps@figure{htbp}
\makeatother

\makeatletter
\@ifpackageloaded{caption}{}{\usepackage{caption}}
\AtBeginDocument{%
\ifdefined\contentsname
  \renewcommand*\contentsname{Table of contents}
\else
  \newcommand\contentsname{Table of contents}
\fi
\ifdefined\listfigurename
  \renewcommand*\listfigurename{List of Figures}
\else
  \newcommand\listfigurename{List of Figures}
\fi
\ifdefined\listtablename
  \renewcommand*\listtablename{List of Tables}
\else
  \newcommand\listtablename{List of Tables}
\fi
\ifdefined\figurename
  \renewcommand*\figurename{Figure}
\else
  \newcommand\figurename{Figure}
\fi
\ifdefined\tablename
  \renewcommand*\tablename{Table}
\else
  \newcommand\tablename{Table}
\fi
}
\@ifpackageloaded{float}{}{\usepackage{float}}
\floatstyle{ruled}
\@ifundefined{c@chapter}{\newfloat{codelisting}{h}{lop}}{\newfloat{codelisting}{h}{lop}[chapter]}
\floatname{codelisting}{Listing}

\makeatother
\makeatletter
\@ifpackageloaded{caption}{}{\usepackage{caption}}
\@ifpackageloaded{subcaption}{}{\usepackage{subcaption}}
\makeatother

\ifLuaTeX
  \usepackage{selnolig}  % disable illegal ligatures
\fi
\usepackage[]{natbib}
\usepackage{bookmark}

\IfFileExists{xurl.sty}{\usepackage{xurl}}{} % add URL line breaks if available
\hypersetup{
  pdftitle={Title},
  pdfauthor={Author 1; Author 2},
  pdfkeywords={3 to 6 keywords, that do not appear in the title},
  colorlinks=true,
  linkcolor={blue},
  filecolor={Maroon},
  citecolor={Blue},
  urlcolor={Blue},
  pdfcreator={LaTeX via pandoc}}

\usepackage[ruled,vlined]{algorithm2e}

\newcommand{\anon}1

\usepackage{amsthm}
\newtheorem{remark}{Remark}
\newtheorem{definition}{Definition}
\newtheorem{lemma}{Lemma}
\newtheorem{proposition}{Proposition}

\usepackage{appendix}

\begin{document}

\def\spacingset#1{\renewcommand{\baselinestretch}%
{#1}\small\normalsize} \spacingset{1}

%%%%%%%%%%%%%%%%%%%%%%%%%%%%%%%%%%%%%%%%%%%%%%%%%%%%%%%%%%%%%%%%%%%%%%%%%%%%%%

\if1\anon
{
  \title{\bf Heterogeneous Responses to Continuous Treatments: A Cluster-Based Causal Framework}
 \author{%
    \parbox{\textwidth}{\centering
      Augusto Cerqua\\
      \emph{Department of Social and Economic Sciences, Sapienza University of Rome, Italy}\\[0.6em]
      Roberta Di Stefano\\
      \emph{Department of Economics, University of Molise, Italy}\\[0.6em]
      Raffaele Mattera\\
      \emph{Department of Mathematics and Physics, University of Campania Luigi Vanvitelli, Italy}
    }%
  }
  \maketitle
} \fi

\if0\anon
{
  \bigskip
  \bigskip
  \bigskip
  \begin{center}
    {\large\bf
    Heterogeneous Responses to Continuous Treatments: A Cluster-Based Causal Framework}
\end{center}
  \medskip
} \fi

\abstract{When treatments are non-randomly assigned, continuous, and yield heterogeneous effects at the same intensity, causal identification becomes particularly challenging. In such contexts, existing approaches often fail to provide policy-relevant estimates of the relationship between treatment intensity and outcomes, especially in the presence of limited common support. To fill this gap, we introduce the Clustered Dose-Response Function (Cl-DRF), a novel estimator designed to uncover the continuous causal relationship between treatment intensity and the dependent variable across distinct subgroups. Our approach leverages both theoretical and data-driven sources of heterogeneity, relying on relaxed versions of the conditional-independence and positivity assumptions that are plausible across various observational settings. We apply the Cl-DRF estimator to estimate subgroup-specific dose–response relationships between European Cohesion Funds and economic growth. In contrast to much of the literature, higher funding increases growth in more developed regions without diminishing returns, while limited absorptive capacity prevents other regions from fully benefiting.}

\noindent%
Continuous treatment,
Heterogeneous treatment effects,
Clustering,
Potential outcome,
European Union funds
\vfill

\maketitle

\section{Introduction}
\noindent Many treatments are non-binary and continuously distributed, making the identification of causal effects a nontrivial problem \citep{athey2017state,fong2018covariate}. Moreover, the complexity of applied empirical contexts typically implies that treatments are allocated based on factors correlated with potential outcomes, violating random assignment and further complicating identification of the causal relationship between treatment intensity and outcomes.

The literature has introduced several estimators specifically designed to address the challenges posed by continuous treatments, significantly broadening the methodological toolkit available to researchers \citep{hirano2004propensity, imai2004causal, kennedy2017non, fong2018covariate, zhao2020propensity, wu2022matching, huling2023independence}. These estimators are designed to recover an unbiased estimate of the causal Average Dose–Response Function (ADRF), a population-based estimand describing how the outcome of the average unit in the population varies with the treatment dosage \citep{tubbicke2023ebct}. Identification relies on two fairly stringent assumptions: weak unconfoundedness and positivity\footnote{Weak unconfoundedness is also referred to as ignorability, no unmeasured confounding, or the conditional independence assumption; the positivity assumption is also referred to as overlap or the common support assumption.}. The first assumption states that, once we account for the observed pre-treatment characteristics, the potential outcomes are independent of the treatment assignment at each value of the treatment intensity, while positivity requires that, for every combination of these characteristics, each unit could, in principle, be observed at any treatment level. Under these assumptions, existing estimators for continuous treatments identify the ADRF, while accommodating treatment effect heterogeneity that is fully captured by observed covariates. 
However, these estimators are not suitable for empirical contexts characterized by treatment effect heterogeneity across units and limited common support across units. Indeed, we show that under these circumstances they fail to consistently recover the true ADRF and, moreover, that the ADRF itself is not the most policy-relevant estimand. 

Treatment effect heterogeneity may arise when the causal effect differs systematically across subpopulations, even when units receive the same treatment intensity. These variations may reflect  differences in baseline characteristics, contextual factors, or differential responses to the treatment. Although the presence of heterogeneous effects is well established in the binary treatment literature \citep{athey2016recursive, wager2018estimation, de2020two, callaway2021difference}, it remains largely overlooked in the continuous treatment literature. %Existing estimators for continuous treatments are built around a population-level ADRF and therefore do not target an estimand that allows for cluster-specific causal relationships. As a result, when treatment effects are heterogeneous across subgroups, these estimators impose a single average dose–response function on the entire population,
Revealing systematic differences in treatment responses necessitates subgroup-specific estimates that are informative about heterogeneous effects and relevant for policy design. %Accounting for such structured heterogeneity requires moving beyond the ADRF framework. Moreover, when heterogeneity is accompanied by group-specific differences in common support, population-level ADRF estimators yield a biased estimates of the dose–response relationship (see Section \ref{Sec:motivating}).

To bridge this methodological gap, we introduce the Clustered Dose Response Function (Cl-DRF) approach. The Cl-DRF characterizes the continuous causal relationship between treatment intensity and outcomes by allowing for multiple dose response functions (DRFs) indexed by clusters, rather than maintaining a single population average ADRF. We first define a novel causal estimand, namely cluster-specific DRFs, and then propose an estimator to recover them. The Cl-DRF framework allows the causal relationship between treatment intensity and outcomes to vary across distinct subgroups or clusters\footnote{Throughout the paper, we use the terms \textit{subgroups} and \textit{clusters} interchangeably.}, each governed by a potentially different underlying mechanism. Following \cite{hirano2004propensity}, we adopt a parametric strategy to model treatment assignment and outcomes. Although less flexible than fully nonparametric alternatives \citep[e.g.,][]{kennedy2017non, colangelo2025double}, this choice provides the tractability needed to jointly estimate the clustering structure and the cluster-specific DRFs.
A key feature of the Cl-DRF estimator is its reliance on assumptions that are plausible across various observational settings. Indeed, while the Cl-DRF estimator adheres to the weak unconfoundedness and positivity framework, it relies on a less stringent version of these assumptions, requiring their validity exclusively within the identified subgroups. For instance, while it is highly unlikely for every unit to maintain a non-zero conditional density across all treatment levels \citep{branson2023causal}, meeting the positivity assumption becomes more plausible within a subgroup characterized by similar covariate values, contextual factors, responses of the outcome to the treatment and, possibly, a smaller range of treatment intensities. By mitigating concerns about assumptions' validity, Cl-DRF can be applied to a broader range of empirical contexts, particularly those characterized by likely heterogeneous treatment effects. 

Since in most empirical applications treatment effect heterogeneity cannot be known \textit{a priori}, the Cl-DRF approach accommodates both theory-driven subgroup definitions, grounded in observed covariates, and data-driven procedures for subgroup identification. When implementing the data-driven version of the approach via the Cl-DRF estimator, we do not assume the number of clusters to be known in advance. Instead, the number of clusters is treated as an unknown parameter and jointly estimated within the Cl-DRF approach. Moreover, we do not assume \textit{a priori} that the likely heterogeneity depends solely on one or more covariates. Instead, we consider that it could stem from the varying response of the outcome to the treatment, which is a function of the covariates. Our focus is on settings where not only the magnitude but also the shape of the dose–response relationship varies across subgroups, in ways that cannot be captured by simple interactions between covariates and treatment intensity. To determine the clusters and estimate the corresponding DRFs, we develop an iterative procedure that alternates between updating cluster membership and estimating cluster-specific treatment assignment models, summarized by the generalized propensity score (GPS), and cluster-specific outcome regressions. This structure mirrors standard clusterwise regression approaches \citep{spath1979algorithmus,demidenko2018next,sugasawa2021grouped}, allowing heterogeneity to emerge from differences in the treatment--outcome relationship across clusters.

To illustrate the properties of the Cl-DRF estimator, we conduct a series of simulations designed to assess its performance under various scenarios. %, comparing its accuracy and robustness against existing approaches.
We next apply the Cl-DRF estimator to quantify the impact of the European Union (EU) Cohesion Policy on economic growth. This setting is particularly suitable for our framework, as EU funds are continuously distributed, non-random, and likely to generate heterogeneous returns depending on regions’ pre-treatment characteristics and absorptive capacity. This empirical application not only showcases the estimator’s ability to handle complex, real-world data but, more importantly, deepens our understanding of the economic effects of EU place-based interventions. The resulting causal estimates enable policymakers to fine-tune the allocation of funds by identifying clusters of regions where marginal returns are highest and those that are unable to benefit from additional support. These insights, in turn, help direct resources toward the most promising areas while mitigating the risk of inefficiently subsidizing regions with limited growth potential. 

The paper structure is as follows. Section \ref{Sec:motivating} illustrates a motivating example with simulated data, while Section \ref{Sec: literature review} discusses previous studies dealing with causal inference for continuous treatment and treatment effects heterogeneity. Section \ref{Sec: methodology} develops the proposed methodology by formalizing the causal framework and the target estimand, introducing the Cl-DRF estimator under its identifying assumptions, and addressing the testing of the cluster structure hypothesis and the selection of the number of clusters. Section \ref{Sec: application} applies the proposed estimator to derive policy-relevant estimates guiding the allocation of European Union (EU) funds assignment to stimulate regional economic growth, while Section \ref{Sec: conclusions} concludes with final remarks.

\section{A motivating example}
\label{Sec:motivating}

\noindent Observational studies in the social sciences often involve continuous treatments, heterogeneous treatment effects, and non-random treatment assignment that induces systematic associations between treatment intensity and observed covariates. These associations can generate severe limitations in common support: for some covariate profiles, it is difficult, both conceptually and empirically, to observe units receiving widely different treatment intensities.\footnote{In other disciplines, this limitation is generally less significant, as it is often easier to identify similar units subjected to widely varying treatment intensities. For example, in medicine, patients hospitalized for a heart attack may share similar characteristics yet exhibit markedly different levels of creatinine, representing the quantitative exposure \citep{austin2019assessing}. Similarly, in environmental health, comparable areas may be affected by pollution at significantly different dosages \citep{wu2022matching}.} Consider, for instance, public subsidies to private firms. Subsidy amounts are continuously distributed and depend on regional and sectoral characteristics, as well as on managerial quality. As a result, the relationship between treatment intensity and covariates may differ across clusters. Moreover, the effect of subsidies on firm outcomes, such as productivity, employment, or innovation, is likely to vary across these clusters, reflecting differences in market conditions and in the quality of investment projects. In such settings, estimating a single ADRF over a wide range of treatment intensities may not be credible. By contrast, it may be feasible to construct theory-based or data-driven clusters, each characterized by a more limited range of treatment intensities. Under these circumstances, cluster-specific DRFs can be estimated. The motivating example below mimics this setting.

Suppose there are four groups of units that differ both in treatment assignment and in the way outcomes respond to treatment. Within each group, treatment intensities lie in a relatively narrow range, common support is limited across groups, and treatment effects may differ in sign and exhibit nonlinearities. If this underlying structure is ignored and a single ADRF is estimated, these heterogeneous response patterns are averaged out. Moreover, the resulting ADRF is likely to be biased due to violations of the positivity assumption. Let $i= 1, \dots, n$ index the observational units in the sample. Let $c=1,\dots,C$ index the clusters, and assume that each unit $i$
belongs to exactly one cluster. Let $\mathbf{y}$ $(y_i, i=1,\dots,n)$ denote the $n$-dimensional vector of observed outcomes. Moreover, let $\mathbf{t}$ $(t_i, i=1,\dots,n)$ be the $n$-dimensional vector of observed treatments, where $t_i \in \mathcal{T} = [t_{\min}, t_{\max}] \subset \mathbb{R}$, and let $\mathbf{X}$ be the $n\times p$ matrix of pre-treatment covariates, where $\mathbf{x}_i$ is the $p$-dimensional vector of covariates associated with each $i$-th unit. We consider a sample size of $n=800$ units that are clustered into $C=4$ equally sized groups and $p=2$ vectors of pre-treatment covariates. For notational convenience, we augment the covariate vector with an intercept, $\tilde{\mathbf{x}}_i = (1, x_{1i}, x_{2i})'$, so that the treatment model
\begin{equation}
    t_{i} \mid \tilde{\mathbf{x}}_i, c \sim N\left(\boldsymbol{\beta}'_c \tilde{\mathbf{x}}_i, 1\right)
\end{equation}
has cluster-specific parameter vectors $\boldsymbol{\beta}_c\in\mathbb{R}^3$.
We set $\boldsymbol{\beta}_1 = [3,1.5,1.2]$, $\boldsymbol{\beta}_2 = [3,1.8,1.5]$, $\boldsymbol{\beta}_3 = [3,2,1.8]$ and $\boldsymbol{\beta}_4 = [3,2.2,2]$. We simulate $x_{1i} \sim U [0, 0.4]$ and $x_{2i} \sim U [0, 0.5]$ for Cluster 1, $x_{1i} \sim U [0.2, 0.6]$ and $x_{2i} \sim U [0.3, 0.6]$ for Cluster 2, $x_{1i} \sim U [0.5, 0.8]$ and $x_{2i} \sim U [0.5, 0.9]$ for Cluster 3, and $x_{1i} \sim U [0.7, 1]$ and $x_{2i} \sim U [0.7, 1]$ for Cluster 4, where $U[a,b]$ denotes the continuous uniform distribution with support $[a,b]$. Then, we simulate the outcome variable as a function of the treatment while allowing for confounding through the covariates,\footnote{In Appendix \ref{motivating_noconf} we model the outcome as a function of the treatment only. Since the treatment depends on covariates, the outcome implicitly depends on covariates as well. This reflects a strategy in which covariate–outcome relationships arise indirectly through the treatment assignment mechanism.} and assume the following within–cluster relationships:
\begin{equation}
\label{simulexample}
y_i=\begin{cases}
5+ 2 t_i+ 1.6 t_i^2 
  + t_i\left(\gamma_{11} x_{1i} + \gamma_{21} x_{2i}\right) + e_{i} & \text { for } c=1,\\[0.25em]
15 - t_{i}-1.6 t_i^2 
  + t_i\left(\gamma_{12} x_{1i} + \gamma_{22} x_{2i}\right) + e_{i} & \text { for } c=2,\\[0.25em]
-5 +2t_i 
  + t_i\left(\gamma_{13} x_{1i} + \gamma_{23} x_{2i}\right) + e_{i} & \text { for } c=3,\\[0.25em]
25 - t_{i} 
  + t_i\left(\gamma_{14} x_{1i} + \gamma_{24} x_{2i}\right) + e_{i} & \text { for } c=4,\\
\end{cases}
\end{equation}
where $e_i \sim N(0,1)$ for all $i$, and $\left(\gamma_{1c}, \gamma_{2c}\right)$ are cluster–specific moderation parameters determining how the covariates $(x_{1i}, x_{2i})$ modify the marginal effect of the treatment within cluster $c$.
We set $(\gamma_{1c})_{c=1}^4 = (0.5,\,0.8,\,1.1,\,1.4)$ and
$(\gamma_{2c})_{c=1}^4 = (0.3,\,0.6,\,0.9,\,1.2)$. 

%This design mimics policy settings where both the assignment mechanism and the effectiveness of the treatment differ across groups, so that estimating a single ADRF blurs substantively different patterns of causal response.
We remark that in this setting we assume two sources of heterogeneity: one in the covariate–treatment relationship and another in the treatment–outcome relationship. The first arises because treatment assignment depends on covariates in a cluster-specific (i.e., non-uniform) way. The second reflects the fact that units in different clusters exhibit heterogeneous responses to the same level of treatment. Both sources of heterogeneity are likely to coexist in practical settings. For example, subsidy amounts may depend on regional and sectoral characteristics as well as managerial quality; consequently, the relationship between treatment intensity and covariates may vary across clusters. Moreover, the effect of subsidies on firm outcomes, such as productivity, employment, or innovation, is likely to differ across these clusters, reflecting variation in market conditions and in the quality of investment projects.

\begin{figure}[!htb]
    \centering
    \includegraphics[width=0.5\linewidth]{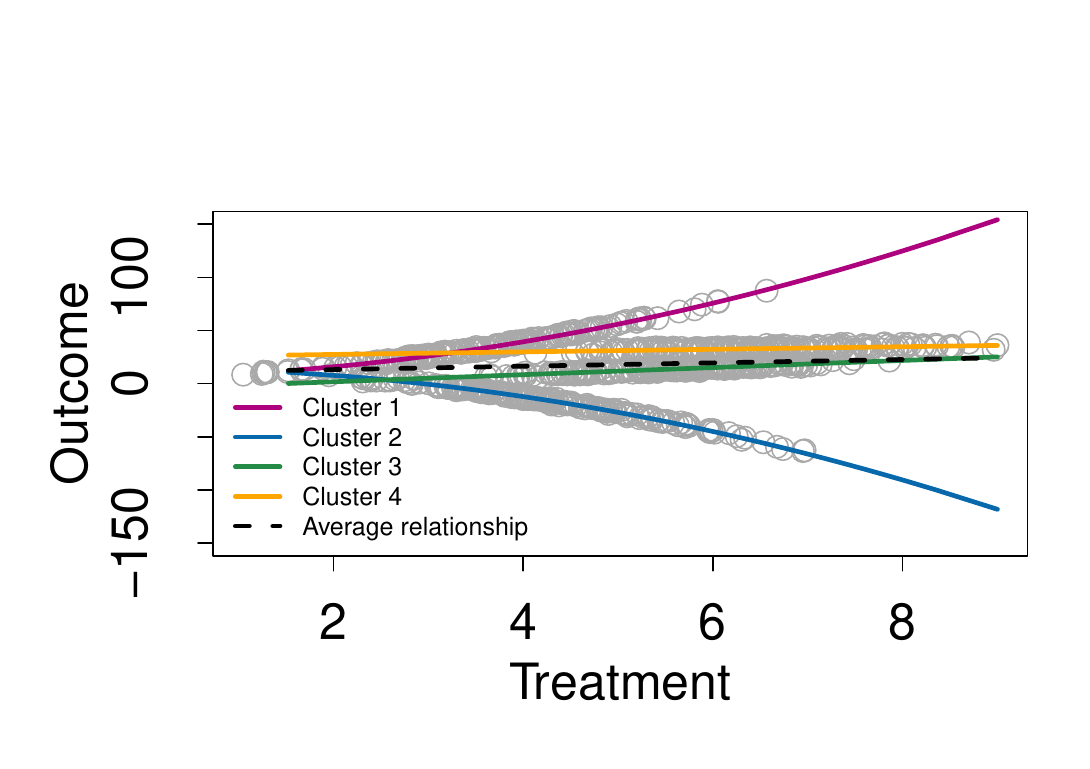}\quad
    \includegraphics[width=0.35\linewidth]{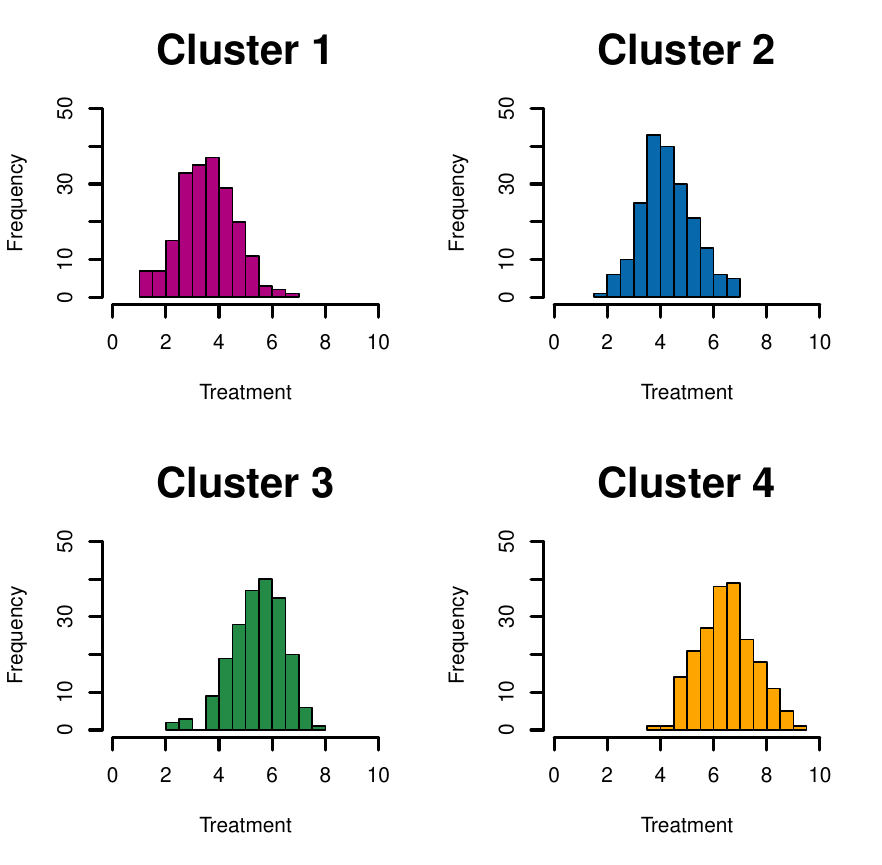} 
    \caption{Relationship between treatment and outcome for all units and at the cluster level (left-hand side) and histogram of distribution of treatment within clusters (right-hand side). Simulated data with $n = 800$ units divided into $C = 4$ equally sized clusters are used. Treatment is assigned based on cluster-specific functions of two pre-treatment covariates, and outcomes are generated using distinct treatment–outcome relationships within each cluster.}
     \label{fig:rel}
\end{figure}

Figure \ref{fig:rel} illustrates the association between treatment and outcome for each cluster. Notice that although the support for treatment varies among clusters, we display the true relationship between outcome and treatment also in areas outside the common support to take into account that theoretically each unit could have received any possible treatment intensity level, i.e., we assume that the positivity condition holds across the entire range of possible treatment intensities. However, due to the cluster-specific assignment mechanism, some treatment values occur with extremely low probability for particular groups, leading to very limited practical overlap in the observed data. 
The dashed black line represents the averaged relationship between outcome and treatment for the average unit (without considering the clustering information), i.e., the population ADRF.

\begin{figure}[!htb]
    \centering  
        \includegraphics[width=0.7\linewidth]{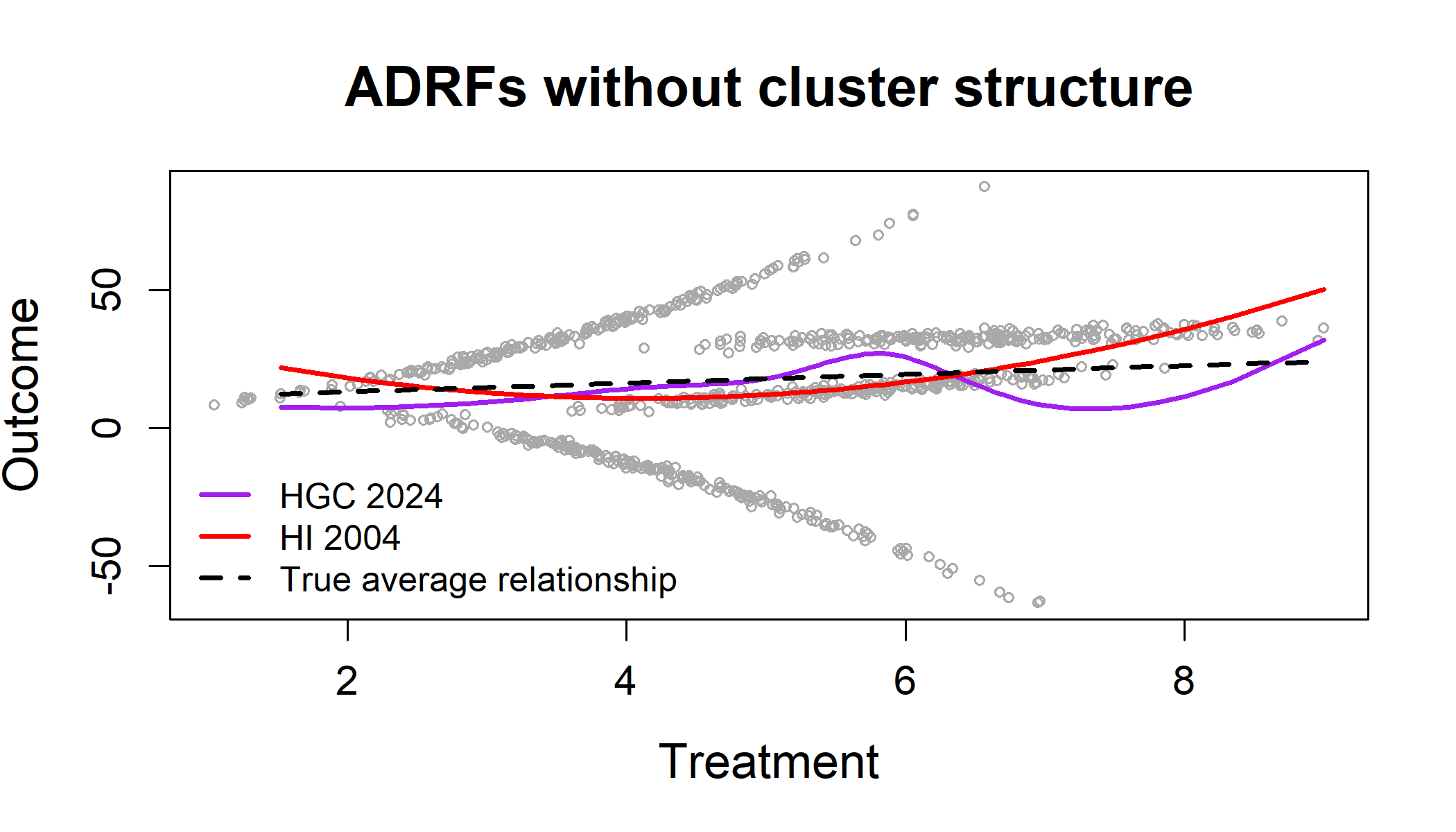}
    \caption{ADRF without cluster structure by using the \cite{hirano2004propensity} approach and following \cite{huling2023independence}. \textit{True average relationship} (dashed black line) represents the average of the four simulated relationships as in equation \ref{simulexample}; \textit{HI2004} (red line) is the estimated ADRF by using the \citep{hirano2004propensity} approach; \textit{HGC 2024} (purple line) is the ADRF estimated by using the \citep{huling2023independence} approach.}
    \label{fig:overall}    
\end{figure}

In this scenario, we argue that estimating a single ADRF for the relationship between treatment intensity and the outcome obscures heterogeneity in the causal link and thus lacks policy relevance. Indeed, this averaging process overlooks the existence of clusters where a strongly positive (or negative) relationship between treatment and outcome is evident. After generating the data, we first estimate the ADRF by using both the parametric  GPS approach of \cite{hirano2004propensity} and the weighted nonparametric approach of \cite{huling2023independence} (see Section \ref{lit: continuous treatment} for further details) to the whole set of 800 units, i.e., without taking into account the presence of clusters. The estimated cluster-based DRFs are displayed in Figure \ref{fig:overall}. 

The ADRF estimated using the \cite{hirano2004propensity} estimator exhibits a convex shape and deviates substantially from the true simulated population ADRF, depicted by the dashed line. Similarly, applying the estimator proposed by \cite{huling2023independence} yields comparable shortcomings. In this case, the estimated ADRF suggests a mildly nonlinear relationship between the outcome and treatment intensity, whereas the true relationship is linear and flat. From a policy perspective, this pooled ADRF would be interpreted as indicating modest returns at low treatment intensities and negligible or even negative effects at higher intensities. However, this pattern is purely an artifact of averaging across groups with sharply heterogeneous responses, with some groups benefiting from higher doses and others not. Therefore, in the setting under analysis, the ADRF is not only an estimand of limited policy relevance but also one that cannot be consistently estimated using existing methods.

\begin{figure}[!htb]
    \centering
\includegraphics[width=0.8\linewidth]{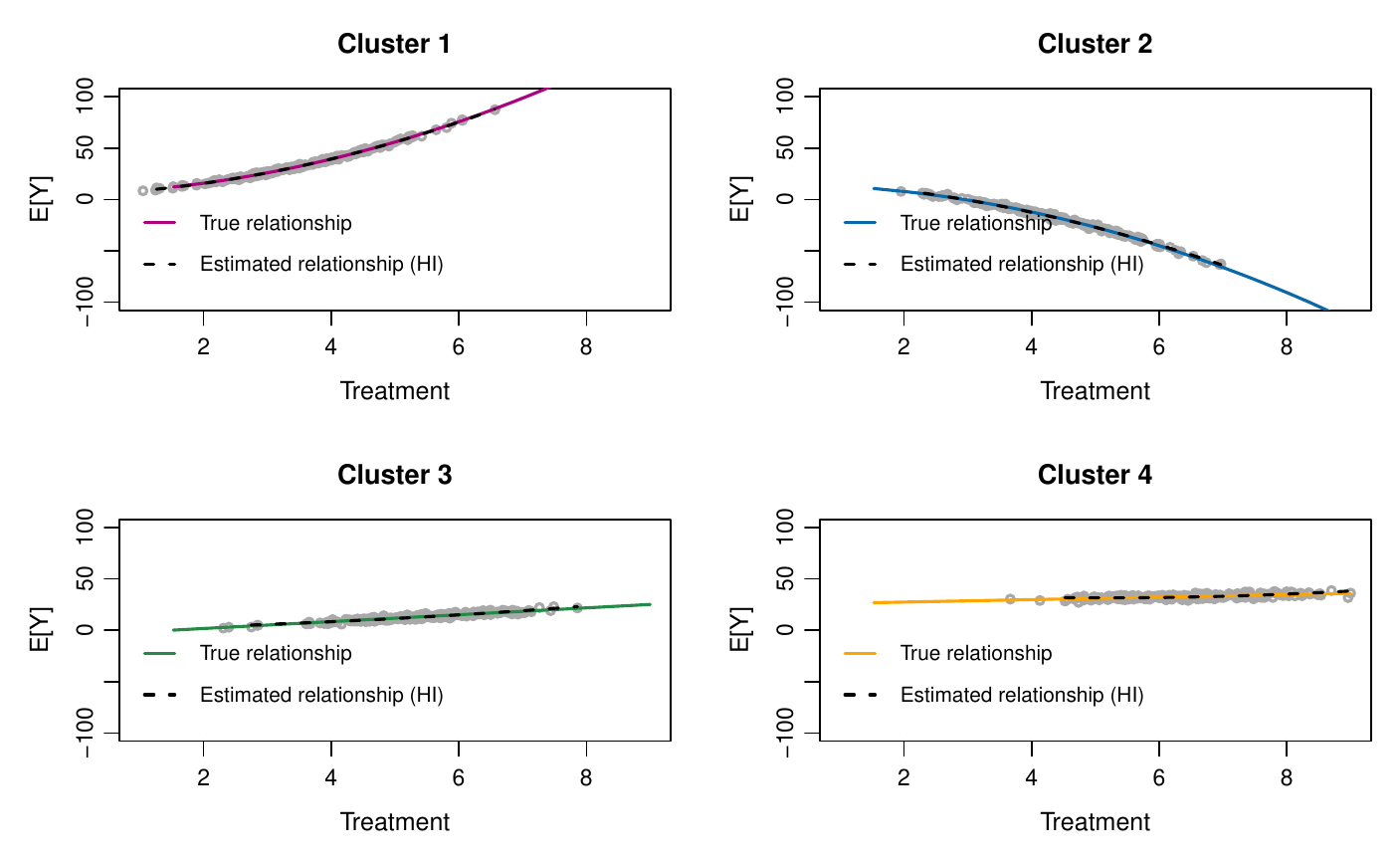} 
\caption{DRFs estimated using the \citep{hirano2004propensity} approach within each cluster, under the assumption that the cluster structure is known, compared to the true simulated relationships. The colored solid lines represent the four simulated relationships for all values of treatment as in equation \ref{simulexample}; the dashed black lines are the estimated DRFs by using the \citep{hirano2004propensity} approach.}
    \label{fig:adrfshirano}    
\end{figure}

Let us assume knowledge of the partition of the units into four groups as in Figure \ref{fig:rel}, that is, we know exactly what units share similar treatment-outcome relationships. Under this assumption, we can apply the \cite{hirano2004propensity} approach within each of these known groups to estimate cluster-specific DRFs. As shown in Figure \ref{fig:adrfshirano}, the estimated DRFs (dashed black lines) perfectly resemble the true relationships (colored lines) within each cluster. The same result holds when using the approach of \cite{huling2023independence} (see Figure \ref{fig:adrfshuling} in Appendix \ref{drf_huling}). This exercise shows that, conditional on knowing the true cluster structure, standard estimators are perfectly capable of recovering the correct DRF within each group. The challenge, therefore, is not the lack of flexible estimators per se, but the absence of a framework that jointly discovers the relevant clusters and estimates their corresponding DRFs. However, policymakers often lack precise knowledge about which groups may exhibit distinct treatment–outcome relationships, meaning that the assumption of known cluster structure is not realistic in practice. This raises the need for a novel modeling framework that contemporaneously identifies the grouping structure of the units in terms of the treatment-outcome relationship and estimates a DRF within each of these groups. The Cl-DRF approach aims to fill this need. In Section \ref{again_motivatingexample}, we apply the Cl-DRF estimator to the same simulated setting without assuming prior knowledge of the cluster structure. We show that the estimator correctly identifies the true number of clusters, assigns units to clusters with high accuracy, and closely recovers the underlying simulated dose–response relationships.% Thus, in the following, we will disregard the information about the cluster structure and the number of clusters. 

\section{Related works}
\label{Sec: literature review}

\noindent In the context of policy evaluation, the simplification of treatment to binary terms and the assumption of homogeneous treatment effects across units have long represented the prevailing evaluation framework, aiding in the tractability and interpretation of empirical findings. However, in most circumstances, these simplifications fail to capture the complex nature of policy interventions, which can vary in intensity (e.g., firms receiving public subsidies are generally financed with different amounts), and produce heterogeneous effects even at a fixed treatment intensity, i.e., heterogeneous responses to a homogeneous treatment (e.g., training programs for the unemployed might assist some individuals in finding a job, while potentially disadvantaging others who might have secured a similar job more quickly without the training).

Addressing these limitations, two parallel strands of literature have emerged. The first line of research—reviewed in Section \ref{lit: continuous treatment}—models the relationship between continuous treatments and outcomes \citep[e.g.,][]{hirano2004propensity, fong2018covariate, wu2022matching} without accounting for the potential heterogeneity of treatment effects. Another strand of literature examines the varied responses to the same treatment across different units \citep[e.g.][]{athey2016recursive, wager2018estimation, de2020two, callaway2021difference} within the binary treatment framework and it is reviewed in Section \ref{lit: heterogeneity}.

\subsection{Approaches for estimating ADRF}
\label{lit: continuous treatment}
\noindent While dichotomizing a continuous treatment variable is known to lead to information loss and potentially compromise data analysis quality \citep{fong2018covariate}, managing the inherent complexity of a continuous treatment framework poses significant challenges, especially in observational studies where the continuous treatment is not randomly assigned. A common approach to reducing confounding by observed variables is to use the propensity score generalized for the setting of continuous treatments \citep{hirano2004propensity}. %\sout{ With this approach, ADRF can be estimated using the weighted regression of the outcome of the treatment, where the weights are proportional to the inverse of the conditional density of the treatment given the covariates, the generalized propensity score (GPS).}
This method aims to provide an unbiased estimate of the ADRF, which relates changes in the outcome variable to changes in treatment for the average unit. Each point of the ADRF represents the expected value of the potential outcome variable at a given treatment level of interest. The estimation process can be split in two steps:
\begin{enumerate}
    \item The \textit{treatment model estimation}, which produces the generalized propensity score (GPS), i.e., the conditional density of being assigned to a particular level of treatment given a set of pre-treatment observed covariates. GPS can be seen as a balancing score.
    \item The \textit{outcome model estimation}, which uses the GPS to correct for treatment endogeneity in the outcome model. Adjusting for the GPS allows units with sufficiently similar characteristics but different treatment intensities to be compared, thus enabling the estimation of potential outcomes at different levels of treatment. By rebalancing the sample over the observed confounders, this approach allows unbiased estimation of the potential outcomes, provided that the assumptions of conditional independence and positivity hold (see below).
    However, the outcome model itself can be specified in various ways, depending on the desired balance between structure and flexibility. Traditional implementations rely on parametric outcome regression or inverse-probability-weighted models \citep{robins2000marginal}, while semiparametric variants employ local regression techniques \citep{flores2012estimating}. More recent contributions focus on doubly robust estimators, which combine outcome and treatment models and deliver both consistency and valid inference when at least one nuisance component is correctly specified \citep{kennedy2017non, colangelo2025double}.
\end{enumerate}

\noindent The difficulty of correctly specifying a conditional distribution is exacerbated by the increased dimension of the pre-treatment covariates to be controlled for, that is the confounders.
To tackle this issue, \cite{wu2022matching} adapt nonparametric matching techniques to the context of continuous treatment. One advantage of this approach is that it clearly separates the design phase from the analysis phase, thereby preserving objectivity because no post-treatment information is used during design. Additionally, as misspecification of a conditional density model can be difficult to diagnose and assess, several studies have focused on directly estimating weights to reduce the correlation between marginal moments of (pre-treatment) covariates and the treatment. Approaches along this line of work include, among others, the generalized covariate balancing propensity score approach \citep{fong2018covariate}, covariate association eliminating weights \citep{yiu2018covariate}, entropy balancing weights \citep{tubbicke2022entropy} and independence weights \citep{huling2023independence}. These approaches prioritize the direct estimation of weights over the explicit estimation and subsequent inversion of a conditional density, and they generally demonstrate superior empirical effectiveness compared to directly modeling the GPS \citep{huling2023independence}, particularly in small samples \citep{cork2025methods}.

All these estimators rely on the conditional independence and the positivity assumptions (see Section \ref{Sec: causal framework} for more details). Given the stringency of these assumptions, the use of such estimators to recover the population ADRF should be restricted to settings in which treatment intensity can be regarded as quasi-randomly assigned, conditional on a set of pre-treatment observable characteristics, and in which each unit can plausibly be exposed to a wide range of treatment intensities.

\subsection{Approaches for estimating heterogeneous treatment effects}
\label{lit: heterogeneity}

\noindent 
It is well-established that units tend to respond differently to binary treatments \citep{athey2016recursive}. This highlights the increasing need for estimators that can assess heterogeneous treatment effects (HTE), focusing not just on average treatment effects but on capturing the diverse responses of different units to a binary treatment. Traditionally, HTE estimation involves repeating the analysis for different subgroups expected to exhibit heterogeneous effects based on prevailing theories (e.g., dividing between young and old individuals or between developed and underdeveloped geographical areas). However, this approach could lead to cherry-picking and multiple testing issues  \citep{athey2016recursive}. In addition, it might miss the identification of the most relevant subgroups of units \citep{wager2018estimation}. Recently, the emphasis has shifted towards data-driven estimation of HTE, with an approach that adopts an algorithm to determine relevant features for estimating treatment effects through the application of machine learning techniques. The goal of this approach is to identify how the effect of a binary treatment varies across a population. These estimators adapt the objective function of well-known machine learning methods, such as decision trees or random forests, with the goal of exploring the heterogeneity of treatment effects relative to a certain set of covariates.

For example, the causal trees introduced by \cite{athey2016recursive} are decision trees tailored to uncover the heterogeneity of treatment effects. In a causal tree, the aim is to achieve the best prediction of the treatment effect and to do this, the algorithm divides the data to minimize the heterogeneity of treatment effects within the leaves (i.e., the differences in potential outcomes), rather than minimizing the heterogeneity of observed outcomes within the leaves. A more complex approach is represented by the Generalized Random Forest (GRF) developed by \cite{athey2019grf}. The GRF iterates over random subsets of the data, constructing a causal tree for each subset. By combining numerous trees, the GRF produces robust and consistent treatment effect estimates under the conditional independence assumption.\footnote{Although GRF can be adapted to continuous‐treatment settings, it estimates average partial effects rather than DRFs and requires untreated units for the analysis.}

In summary, both strands of the literature have produced effective estimators within their respective domains. However, they fall short in observational settings with continuous treatments and heterogeneous unit effects, even at the same treatment intensity. Applying these estimators in such contexts could lead to oversimplified policy recommendations, obscuring the complex impacts of policy measures. In the following section, we introduce the Cl-DRF estimator, a method expressly developed for this evaluation context.

\section{The Clustered Dose-Response-Function Estimator}
\label{Sec: methodology}
\subsection{The causal framework for continuous treatments}
\label{Sec: causal framework}

\begin{figure}[!htb]
     \centering   
     \includegraphics[width=0.6\linewidth]{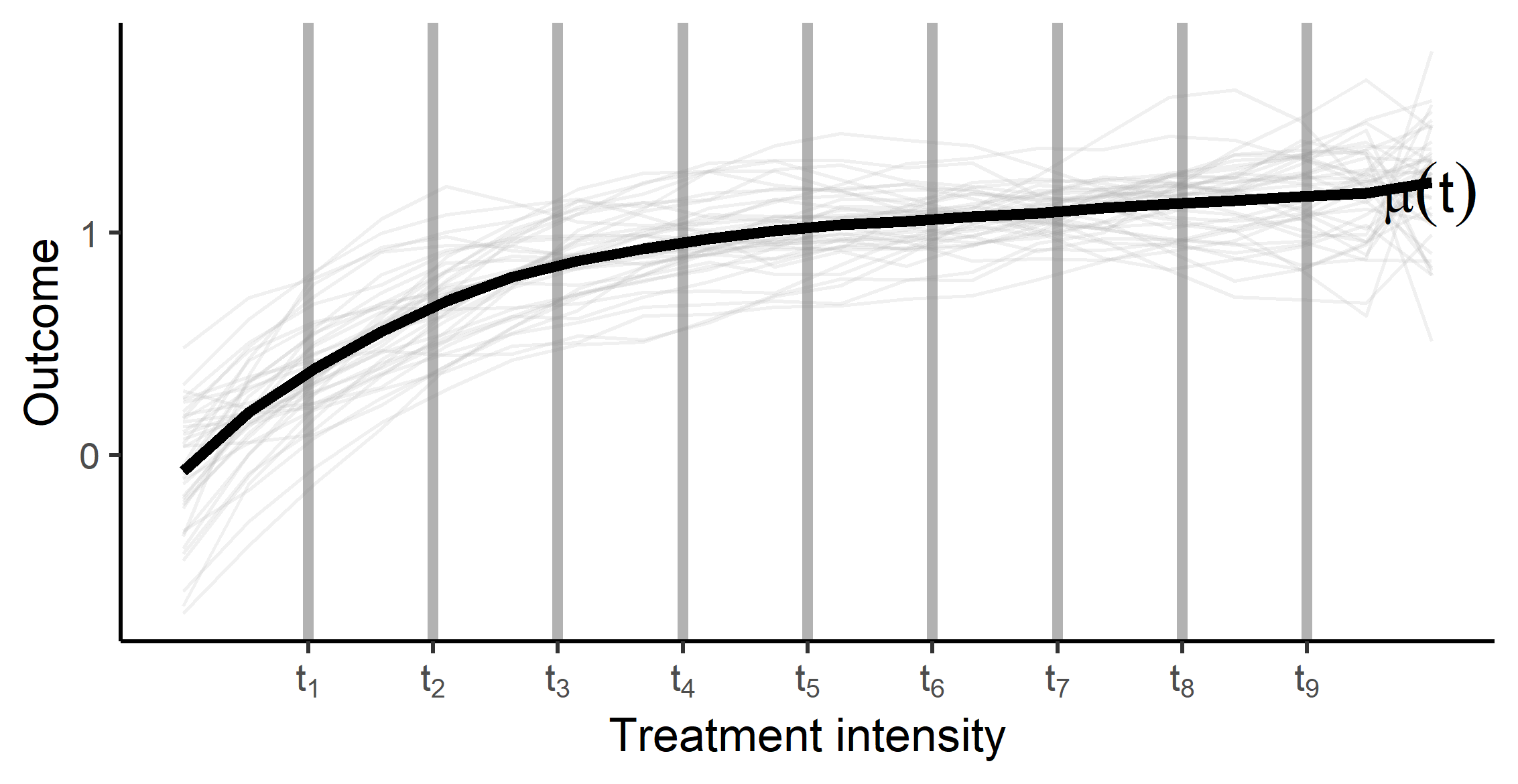}
     \caption{Average dose-response function. Example based on 40 simulated unit-dose responses.}
     \label{fig: adrf}
 \end{figure}

\noindent The potential outcomes model for observational studies \citep{rubin1974estimating} is typically employed in binary contexts. Let $t_i\in \{0,1\}$ be the binary treatment variable for unit $i$, and $y_i$ be the outcome variable for unit $i$. Here, $y_i(1)$ represents the potential outcome under treatment, while $y_i(0)$ denotes the potential outcome in the absence of treatment. Consequently, $y_i(1) - y_i(0)$ can be defined as the unit-level causal effect of a binary treatment. In a more general scenario with continuous treatment, $t_i$ assumes values within a real interval $\mathcal{T} = [t_{\min}, t_{\max}]$. Therefore, $y_i(t)$, where $t \in \mathcal{T}$, signifies the outcomes for unit $i$ when exposed to treatment level $t$, varying within the set $\mathcal{T}$. This is often referred to as the unit-dose response \citep{tubbicke2022entropy}.  So, in a continuous setting, we are interested in investigating the trajectory illustrating how $y_i(t)$ changes across relevant values of $t$. Usually, we are interested in population causal effects, so we consider the expected potential outcomes,
\begin{equation}\label{ardf}
    \mu(t)=\mathbb{E}[y_i(t)],
\end{equation}
also called the average dose-response function (ADRF). Figure \ref{fig: adrf} shows the representation of the potential outcomes path for each unit with grey lines in a simulated example in which we know the actual dose-response function of each unit.\footnote{In practice, we do not observe the potential outcomes path for each unit but only the single value $y_i(t_i)$ corresponding to the observed outcome $y_i$ at the realized treatment intensity $t_i$.} The vertical lines at a generic level $t$ allow us to approximate the ADRF for each level $t$. 
Thus, the population ADRF is the solid black line.
  
\noindent Under random assignment of treatment intensity, differences in average outcomes across treatment levels identify the population ADRF. In contrast, most empirical analyses in the social sciences rely on observational data, where treatment is not randomly assigned. In such studies, treatment intensity typically depends on a set of observed covariates, so careful adjustment for confounding is essential. Therefore, we define the vector of observed pre-treatment covariates for each unit $i$ as $X_i$.  
With observational studies, we need to rely on some strong assumptions to obtain consistent estimates of the ADRF.
Most available estimators for the ADRF in observational settings are based on the same set of assumptions. We refer to A1--A3 as the classical assumptions for ADRF identification.

\medskip
\noindent \textit{Assumption A1 (Consistency).}  
For each unit $i$ and treatment level $t$, if $T_i = t$ then $Y_i = Y_i(t)$. 

\medskip 
\noindent According to the \textit{consistency assumption} the potential outcome of each unit depends solely on the treatment level it receives. In particular, this assumption implies that potential outcomes do not depend on the treatment assigned to other units, that is, there is no interference \citep{cox1958planning}, nor on how the treatment is delivered, provided the treatment level is the same, that is, there are no multiple versions of the treatment. This set of conditions is commonly referred to as the Stable Unit Treatment Value Assumption (SUTVA, \citealt{rubin1980}).

\medskip
\noindent \textit{Assumption A2 (Weak unconfoundedness).}  
The \textit{weak unconfoundedness assumption}\footnote{Unlike the unconfoundedness assumption, weak unconfoundedness does not require the joint independence of all potential outcomes; instead, it requires conditional independence to hold at given treatment levels \citep{becker2012too}. However, in the context under analysis, the difference between the weak unconfoundedness and the unconfoundedness assumptions is more formal than empirically relevant.} implies that $Y_i(t) \perp T_i \mid X_i$, i.e., potential outcomes are independent of treatment conditional on covariates.

\medskip
\noindent The weak unconfoundedness assumption states that once we control for 
observed covariates, how much treatment a unit receives provides no additional information about what its outcome would be at any specific dose. As argued by \cite{bia2012assessing}, such assumption is not testable, difficult to satisfy, and not generally applicable. 

\medskip
\noindent  \textit{Assumption A3 (Positivity).}  For every value of $X_i$, the conditional density of
receiving any treatment level $t \in \mathcal{T}$ is bounded away from zero,
\[
f_{T \mid X}(t \mid X_i) \geq v > 0 .
\]

\medskip
\noindent Also the positivity assumption is crucial for the estimation of causal effects because it guarantees that there is sufficient overlap in the covariate distributions across different levels of the treatment, allowing for meaningful comparisons and generalization of causal effects across the population. In other words, the positivity assumption requires that, for every combination of covariates observed in the data, the conditional density of receiving any treatment level of interest is strictly positive.
Such an assumption is arguably less tenable when the treatment is continuous because it is unlikely that every unit has a non-zero conditional density at every treatment value \citep{branson2023causal}. 
Violations of the positivity assumption arise when certain treatment levels are rarely or never observed within specific regions of the covariate space. This lack of overlap can lead to extrapolation and a biased estimate of the ADRF. In observational studies, researchers must be particularly vigilant for areas where overlap might be insufficient.  \citet{branson2023causal} introduce a framework for trimming and smoothing techniques that  allows for robust estimation even in the presence of limited overlap.\footnote{However, even this approach cannot fully compensate for fundamental violations of the positivity assumption. Furthermore, even if positivity holds, propensity scores close to zero adversely affect the bias and variance of most estimators \citep{li2018balancing}. See also \cite{zhang2025doubly}, who show that propensity score weighting and doubly robust estimators can be inconsistent in the absence of positivity, even when the average dose–response function remains identifiable under strong structural assumptions on the potential outcomes.
} Moreover, recent works have explored alternative estimands and novel identification strategies that circumvent the need for strict positivity by trading off either the scope of the causal estimand or the strength of the underlying identifying assumptions.\footnote{In particular, \cite{schindl2024incremental} introduce incremental effects, an estimand defined under stochastic policy interventions that marginally tilt the observed treatment distribution while preserving its support. 
While a dose–response function characterizes how outcomes vary across treatment intensities, incremental effects collapse this information into a single average effect of a support-preserving policy shift, abstracting from dose-specific heterogeneity in the identified estimand.
Complementarily, \cite{zhang2024nonparametric} retain the ADRF as the target estimand and show that it can be identified without global positivity by leveraging local derivative information, at the cost of imposing stronger structural assumptions on the confounding process.}

\subsection{The Clustered Dose-Response Function approach}

\noindent As shown above, conventional approaches to continuous treatments adopt a population-level identification strategy and thus obscure heterogeneity in treatment responses. We propose an alternative strategy based on local identification within causal regimes, yielding empirically credible and policy-relevant dose–response relationships.

Specifically, if the sample is partitioned into $C \geq 2$ clusters, either based on policy considerations, such as firm size categories (small, medium and large), or using a data-driven procedure, the target estimands become $C$ cluster-specific DRFs. These estimands remain policy-relevant and identifiable under weaker conditions, as weak unconfoundedness and positivity are only required to hold within clusters rather than in the full population (see the remainder of this section). 
More precisely, we target a cluster-specific DRF, which
constitutes the primary causal estimand of the proposed framework.

\begin{definition}[Cluster-specific DRF]
\label{def:cl_drf}
Let $T_i \in \mathcal{T}$ denote a continuous treatment, $Y_i(t)$ the potential
outcome corresponding to treatment level $t$, and
$C_i \in \{1,\dots,C\}$ a cluster indicator indexing subpopulations
characterized by distinct treatment--outcome relationships.
The cluster-specific dose--response function (Cl-DRF) is defined as
\begin{equation}
\mu_c(t) := \mathbb{E}\!\left[ Y_i(t) \mid C_i = c \right],
\qquad c = 1,\dots,C.
\end{equation}
\end{definition}

\noindent This definition makes explicit that the estimand of interest is a
subpopulation-specific causal response, rather than a population-average
quantity. The Cl-DRF therefore allows treatment effects to vary systematically across
clusters, while retaining a clear causal interpretation within each
cluster.

\begin{remark}
The population average dose--response function $\mu(t)$ coincides with the
cluster-specific dose--response functions $\mu_c(t)$ for all $c$ only in the
absence of regime-dependent causal effects.
A sufficient condition for this to occur is that the potential outcome $Y_i(t)$
does not depend on cluster membership, either directly or through effect
moderation.
In such cases, $\mu_c(t)=\mu(t)$ for all $c$ and $t$, and cluster-specific
analysis provides no additional information.
\end{remark}
\noindent Identification and estimation are conducted at the cluster level under the assumptions A1 and B1–B3.

\medskip
\noindent\textit{Assumption B1 (Existence of causal regimes).}  
The population can be partitioned into $C\geq 2$ clusters indexed by
$C_i\in\{1,\dots,C\}$ such that a cluster-specific dose--response function
$\mu_c(t)=\mathbb E[Y_i(t)\mid C_i=c]$ exists and is well defined for all
$t\in\mathcal T_c$. Here, $\mathcal T_c$ denotes the support of the treatment distribution within
cluster $c$.

\medskip
\noindent  \textit{Assumption B2 (Weak unconfoundedness within clusters).}  
For each cluster $c \in \{1,\dots,C\}$ and for all $t \in \mathcal{T}_c$,
\[
Y_i(t) \perp T_i \mid X_i,\, C_i = c .
\]
This condition allows treatment assignment to be confounded at the population
level, while requiring conditional independence to hold locally within each causal regime.

\medskip
\noindent  \textit{Assumption B3 (Positivity within clusters).}  
This assumption rules out extrapolation within clusters and is substantially weaker than requiring overlap over the entire population. For every value of $X_i$ and for each cluster $c$, the conditional density of
receiving any treatment level $t \in \mathcal T_c$ is uniformly
bounded away from zero,
\[
f_{T \mid X, C}(t \mid X_i, C_i = c) \geq v > 0 .
\]

\medskip
\noindent  We remark that Assumptions B2--B3 can be seen as the cluster-specific counterpart of Assumptions A2--A3. Under the assumptions A1 and B1--B3, the cluster-specific dose--response function is
identified. In particular, Assumption~B3 ensures that each cluster-specific dose--response function
$\mu_c(t)$ is identified only over the treatment support actually observed
within cluster $c$, thereby ruling out extrapolation across heterogeneous
regimes. The following lemma establishes identification.

\begin{lemma}[Identification of the cluster-specific dose--response function]
\label{lem:gformula}
Fix $c\in\{1,\dots,C\}$ and $t\in\mathcal T_c$. Under Assumption~A1 and Assumption B1--B3, the
cluster-specific dose--response function
\[
\mu_c(t):=\mathbb E\!\left[ Y_i(t)\mid C_i=c\right]
\]
is identified and admits the representation
\begin{equation}
\label{eq:gformula_clust}
\mu_c(t)
=
\mathbb{E}\!\left[
\mathbb{E}\!\left(
Y_i \mid T_i=t,\, X_i,\, C_i=c
\right)
\mid C_i=c
\right].
\end{equation}
\end{lemma}

\noindent Proof of Lemma \ref{lem:gformula} is shown in Appendix \ref{app:lemma1}. While Lemma~\ref{lem:gformula} provides a valid identification result, direct
conditioning on the full covariate vector $X_i$ is not convenient for causal
adjustment in continuous-treatment settings.
In particular, identification of DRFs requires comparing
units at the same treatment level while ensuring covariate balance.
To this end, \cite{hirano2004propensity} introduce the generalized propensity score (GPS)
as a balancing score that summarizes all confounding information relevant for
treatment assignment at each treatment level.

\begin{definition}[Cluster-specific generalized propensity score]
\label{def:gps}
For each cluster $c$, the GPS is defined as the
conditional density of the treatment given pre-treatment covariates and
cluster membership,
\[
R_{i,c}(t) := f_{T \mid X, C}(t \mid X_i, C_i = c),
\qquad t \in \mathcal{T}_c.
\]
\end{definition}

\noindent The cluster-specific GPS summarizes all confounding information contained in the covariates into a scalar function of the treatment level, conditional on cluster membership. This is the natural restriction of the GPS to cluster $c$. The usefulness of the cluster-specific GPS is based on its balancing property within cluster. Indeed, it directly follows from \cite{hirano2004propensity} that, within each cluster, conditioning on the GPS
suffices to control for confounding in the treatment assignment.

\begin{lemma}[GPS-based representation within clusters]
\label{cor:gps_representation}
Fix $c\in\{1,\dots,C\}$ and $t\in\mathcal T_c$. Under Assumption A1 and Assumptions B1--B3,
\[
\mu_c(t)=\mathbb E\!\left[Y_i(t)\mid C_i=c\right]
=
\mathbb{E}\!\left[
\mathbb{E}\!\left(
Y_i \mid T_i=t,\, R_{i,c}(t),\, C_i=c
\right)
\mid C_i=c
\right],
\]
where $R_{i,c}(t)=f_{T\mid X,C}(t\mid X_i, C_i=c)$ is the cluster-specific
GPS in Definition~\ref{def:gps}.
\end{lemma}

\noindent Proof of Lemma \ref{cor:gps_representation} is shown in Appendix \ref{app:coro1}. This representation provides the basis for estimation strategies that condition
on the treatment level and the cluster-specific GPS, rather than on the full set
of covariates.

\color{black}

\subsection{The Cl-DRF estimation algorithm}
\label{Sec: Cl-DRF algorithm}

\noindent Estimation of the
cluster-specific DRFs requires specifying how cluster
membership is handled in practice.
Two conceptually distinct approaches are in principle available:
\begin{enumerate}%[label=\alph*]
    \item \textit{Theory-based approach.}
    In this case, the sample is divided into $C$ clusters based on the specific features of the treatment under analysis or theoretical predictions, and then a DRF is estimated within each cluster. It is assumed that the researcher knows the relevant clustering structure, i.e., which units belong to which clusters. For instance, \cite{bia2012assessing} study the impact on employment of the amount of financial aid attributed to enterprises analyzing small-sized firms and medium- or large-sized firms separately. Various estimators from the existing literature can be employed to estimate the cluster-specific DRFs. However, conducting separate analyses across multiple pre-defined clusters may exacerbate multiple-testing concerns and introduce scope for selective inference.
    \vspace{0.5pt}
    \item \textit{Data-driven approach.}    
    In this case, the $C$ clusters are not pre-determined, and an algorithm is used to identify clusters and then estimate clustered DRFs. In this setting, the researcher is not required to know the clustering structure ex ante, but instead relies on sufficient signal in the data to recover it.
\end{enumerate}

\noindent We focus on the latter case by employing a cluster-wise extension of the GPS approach of \cite{hirano2004propensity}\footnote{We recall that the idea behind the GPS approach by \cite{hirano2004propensity} is to predict missing potential outcomes at specific values of $t$ by fitting a parametric model of the outcome that includes both the treatment and the GPS as covariates, where the GPS is the conditional density of the treatment given a set of covariates.},
building on ideas from grouped and clustered regression models
\citep[e.g.][]{demidenko2018next,sugasawa2021grouped}. 
This procedure allows us to simultaneously estimate cluster membership
and cluster-specific DRFs, and will be referred to as the
Cl-DRF estimator.
In the Cl-DRF framework, units assigned to the same cluster share common
parameters governing the treatment--outcome relationship, while clusters are recovered through an iterative procedure. 

Following \cite{hirano2004propensity}, we posit a parametric working model for
the treatment assignment mechanism.
However, cluster membership is unknown and
the treatment assignment model is allowed to vary across clusters.
Specifically, we assume a cluster-specific Gaussian specification,
\begin{equation}
T_i \mid X_i, C_i = c \sim \mathcal N\!\left(\boldsymbol{\beta}_c'X_i, \sigma_c^2\right),
\label{eq:gps_clust}
\end{equation}
which induces a cluster-specific GPS.
Evaluated at the observed treatment level, the GPS is
given by
\begin{equation}
R_{i,c}
=
f_{T\mid X,C}(T_i \mid X_i, C_i=c)
=
\frac{1}{\sqrt{2\pi\sigma_c^2}}
\exp\!\left(
-\frac{(T_i - \boldsymbol{\beta}_c'X_i)^2}{2\sigma_c^2}
\right).
\label{eq:gpsclust}
\end{equation}
In the second stage, we model the conditional expectation of $Y_i$ given $T_i$
and the GPS.
Specifically, we posit a parametric working model for
\begin{equation}
\mathbb{E}[Y_i \mid T_i = t, R_{i,c} = r, C_i = c]
= f(t,r;\boldsymbol{\alpha}_c),
\end{equation}
where the functional form $f(\cdot)$ is common across clusters, while the
parameter vector $\boldsymbol{\alpha}_c$ is cluster-specific.
Without loss of generality, we can consider a linear model,
\begin{equation}
f(t,r;\boldsymbol{\alpha}_c)
=
\alpha_{0,c} + \alpha_{1,c} t + \alpha_{2,c} r,
\label{eq:drfclust1}
\end{equation}
though higher-order terms and interactions can be incorporated without altering
the structure of the estimator. Because the GPS depends on cluster membership, the
regressors entering the outcome regression are cluster-specific.

Motivated by the identification result in Lemma~\ref{cor:gps_representation}, we model the conditional expectation of
the outcome given the treatment level and the cluster-specific GPS.
Within each cluster, conditioning on $(T_i,R_{i,c})$ is sufficient for causal
adjustment, and the resulting regression can therefore be used to estimate the
cluster-specific DRFs. In particular, we remark that, within each cluster, the GPS
plays the same role as in the standard framework of
\citet{hirano2004propensity}, thereby justifying this regression-based
estimation strategy. However, the parameters
$\boldsymbol{\alpha}_c$ could be estimated by ordinary least squares conditional on cluster membership and on the GPS
implied by the treatment assignment model.
Indeed, in the present setting cluster membership is unknown and the outcome
regression is allowed to vary across clusters.
As a result, estimation does not reduce to a single linear regression, but
instead takes the form of a cluster-wise linear regression problem \citep[see also][]{spath1979algorithmus,hathaway1993switching,park2017algorithms}. 

To formalize this structure, let $\kappa_{i,c}$ denote cluster membership
indicators, with $\kappa_{i,c}=1$ if unit $i$ belongs to cluster $c$ and
$\kappa_{i,c}=0$ otherwise.
Given a partition of the sample, estimation of the outcome regression parameters
reduces to least-squares fitting within each cluster.
When cluster membership is unknown, however, the indicators
$\{\kappa_{i,c}\}$ constitute additional quantities that must be inferred
from the data. Because the outcome regression parameters and the cluster indicators are not
jointly identified in closed form under this specification, estimation proceeds
by minimizing the following  objective function
\begin{equation}
J:\min_{\{\boldsymbol{\alpha}_c,\kappa_{i,c}\}}
\sum_{i=1}^n \sum_{c=1}^C
\kappa_{i,c}\bigl(Y_i - \boldsymbol{\alpha}_c'Z_i\bigr)^2,
\label{eq:objfunc}
\end{equation}
subject to $\sum_{c=1}^C \kappa_{i,c} = 1$ for all $i$. This formulation naturally leads to an iterative estimation procedure that
alternates between updating the outcome regression parameters and updating
cluster membership. The resulting block-wise updates admit closed-form expressions, which are
summarized in the following proposition.

\begin{proposition}[Closed-form updates in the Cl-DRF algorithm]
\label{prop:updates}
Let $\mathbf{Z} = [\mathbf{z}_1,\dots,\mathbf{z}_n]'$ denote the matrix of
regressors constructed from $(1,T_i,R_i)'$ (thus, without loss of generality, also including their interactions and/or polynomials), where $R_i$ depends on the current
cluster assignment, and let $\boldsymbol{\kappa}_c =
\mathrm{diag}(\kappa_{1,c},\dots,\kappa_{n,c})$ be the diagonal matrix of cluster
membership indicators.
Conditional on $\boldsymbol{\kappa}_c$ and on $\mathbf{Z}$, the minimizer of
\eqref{eq:objfunc} with respect to $\boldsymbol{\alpha}_c$ is given by
\begin{equation}
\boldsymbol{\alpha}_c
=
\left(\mathbf{Z}'\boldsymbol{\kappa}_c \mathbf{Z}\right)^{-1}
\mathbf{Z}' \boldsymbol{\kappa}_c \mathbf{y}.
\label{eq:alpha}
\end{equation}
Conditional on $\{\boldsymbol{\alpha}_c\}_{c=1}^C$, the optimal cluster
assignment satisfies
\begin{equation}
\kappa_{i,c}
=
\mathbf{1}
\left[
\left(y_i - \boldsymbol{\alpha}_c' \boldsymbol{z}_i\right)^2
\le
\left(y_i - \boldsymbol{\alpha}_{c'}' \boldsymbol{z}_i\right)^2,
\ \forall c' \neq c
\right].
\label{eq:kappa}
\end{equation}
Finally,conditional on $\boldsymbol{\kappa}_c$, the GPS
for unit $i$ is updated according to
\begin{equation}
\widehat{R}_{i}
=
\sum_{c=1}^C
\kappa_{i,c}
\frac{1}{\sqrt{2\pi\widehat{\sigma}_c^2}}
\exp\!\left(
-\frac{(T_i-\widehat{\boldsymbol{\beta}}_c'X_i)^2}{2\widehat{\sigma}_c^2}
\right),
\label{eq:gps_update}
\end{equation}
where $(\widehat{\boldsymbol{\beta}}_c,\widehat{\sigma}_c^2)$ are the
maximum-likelihood estimators of the cluster-specific treatment assignment
model parameters.
\end{proposition}

\noindent
Taken together, Proposition~\ref{prop:updates} characterises a block-coordinate
descent procedure in which outcome regression parameters, cluster membership,
and the GPS are updated sequentially until convergence. The proof of Proposition \ref{prop:updates} is shown in Appendix \ref{app:prop1}.

\begin{remark}[Absence of the GPS update step]
\label{rem:no_gps_update}
We are assuming cluster-level heterogeneity for both treatment and outcome models. However, if the treatment assignment mechanism is assumed to be common across clusters,
the GPS does not depend on cluster membership.
In this case, the treatment assignment model is estimated once using the full
sample, and the corresponding GPS remains fixed
throughout the algorithm.
As a result, the third update step in Proposition~\ref{prop:updates} is omitted,
and estimation reduces to an iterative procedure alternating only between
cluster-wise outcome regression and cluster reassignment. This simplified algorithm remains coherent with the cluster-specific
dose--response estimand, provided that the weak unconfoundedness assumption
holds conditional on cluster membership. When the treatment assignment model is assumed to be common across clusters,
the GPS is still defined conditional on $C_i$, but its
functional form does not vary across regimes.
In this case, allowing the treatment assignment model to be common across
clusters affects only the estimation strategy, not the definition or
identification of the Cl-DRF.
\end{remark}

\noindent Taken together, Proposition~\ref{prop:updates} and Remark~\ref{rem:no_gps_update}
define an iterative estimation procedure that belongs to the broad class of
$k$-means--type and grouped regression algorithms.
Theoretical properties of such procedures, including consistency of the
estimated partition and of the cluster-specific regression parameters, have
been extensively studied in the literature under suitable regularity
conditions \citep[e.g.,][]{pollard1981strong,pollard1982central,bonhomme2015grouped,sugasawa2021grouped}. We anticipate that the simulation study reported in detail in Appendix~\ref{app:simulations}
provides strong evidence that the proposed algorithm accurately
recovers the cluster structure and the corresponding cluster-specific
dose--response functions.

\color{black}
\subsection{Choosing the number of clusters}
\label{section:test}

\noindent The Cl-DRF estimator builds on the $k$-means regression methodology and therefore it requires specifying the number of clusters $C$ in advance. To this end, we adopt a penalized empirical risk criterion inspired by
information criteria commonly used in grouped regression models
\citep[e.g.,][]{sugasawa2021grouped}.
For a given number of clusters $C$, we define
\begin{equation}
\label{eq:bic}
\operatorname{IC}(C)
=
\sum_{i=1}^n
\left(y_i - \widehat{\boldsymbol{\alpha}}_{c_i}' \mathbf{z}_i\right)^2
+
\zeta_n \sum_{c=1}^C \operatorname{dim}(\widehat{\boldsymbol{\alpha}}_c),
\end{equation}
where $c_i$ denotes the cluster assignment of unit $i$
and $\operatorname{dim}(\widehat{\boldsymbol{\alpha}}_c)$ denotes the number
of parameters in the cluster-specific regression.
Throughout, we set $\zeta_n = \log n$. We notice that the criterion \eqref{eq:bic} corresponds to a penalized empirical risk based on the
within-cluster sum of squared residuals of the outcome model.
The penalty term controls model complexity through the number of clusters and the dimensionality of the cluster-specific regression functions.
This construction is analogous to information criteria used in
grouped and $k$-means regression settings
\citep[e.g.,][]{sugasawa2021grouped}, and is therefore referred to as
BIC-like.

The criterion \eqref{eq:bic} is evaluated for $C = 1,\dots,C_{\max}$,
where $C=1$ corresponds to a baseline model without clustering.
Since the objective function typically decreases monotonically with $C$,
we select the number of clusters using an elbow-type rule.
Specifically, we identify the value of $C$ that maximizes the vertical
distance between $\operatorname{IC}(C)$ and the straight line connecting
$\operatorname{IC}(1)$ and $\operatorname{IC}(C_{\max})$.
This procedure captures the point at which further increases in $C$
yield only marginal improvements in fit relative to the increase in model
complexity. We emphasize that this selection rule is heuristic in nature and is not
claimed to be consistent for the true number of clusters.
Nevertheless, the simulation results in Appendix \ref{app:simulations}
show that it performs very well in finite samples and accurately recovers
the underlying cluster structure across a wide range of scenarios.

\subsection{Revisiting the motivating example}
\label{again_motivatingexample}
\begin{figure}[!htb]
    \centering
            \includegraphics[width=0.5\linewidth]{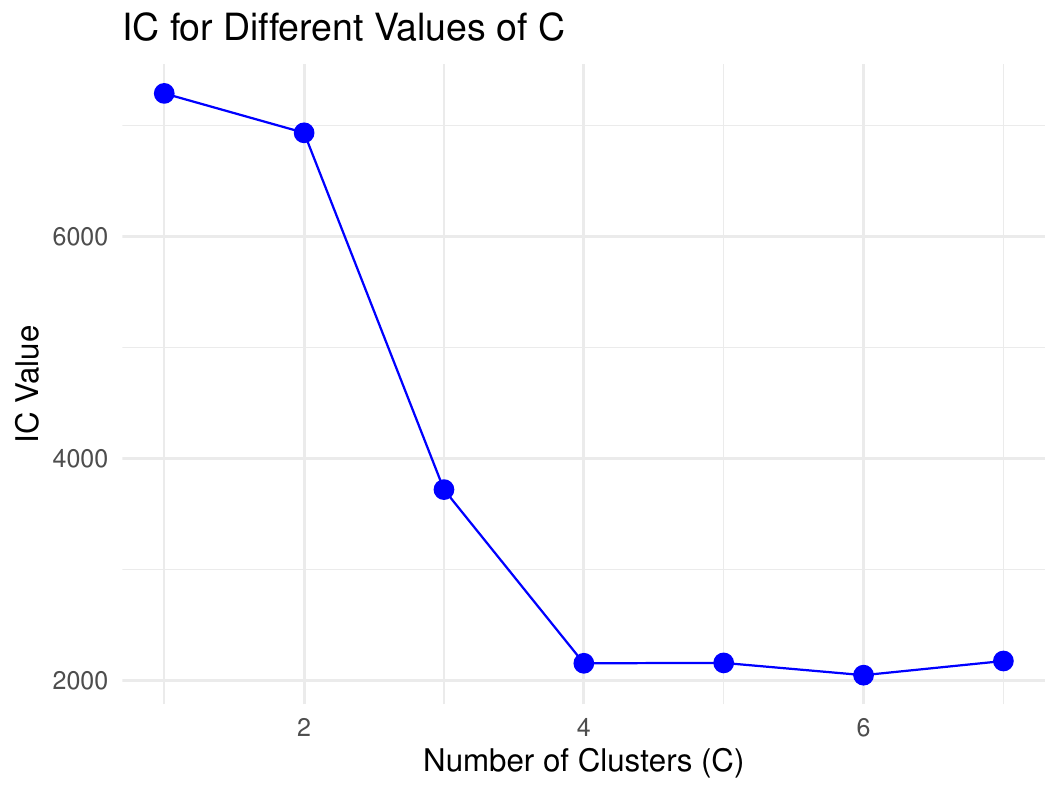}
        \caption{Selection of the number of clusters. IC values \eqref{eq:bic} for different values of $C$, in the set $C = \{1, 2,3,4,5,6,7\}$. We choose $C$ with the Elbow criterion. We detect an elbow for $C=4$.}         
    \label{fig:bicex}
\end{figure}

\begin{figure}[!htb]
    \centering  
\includegraphics[width=0.8\linewidth]{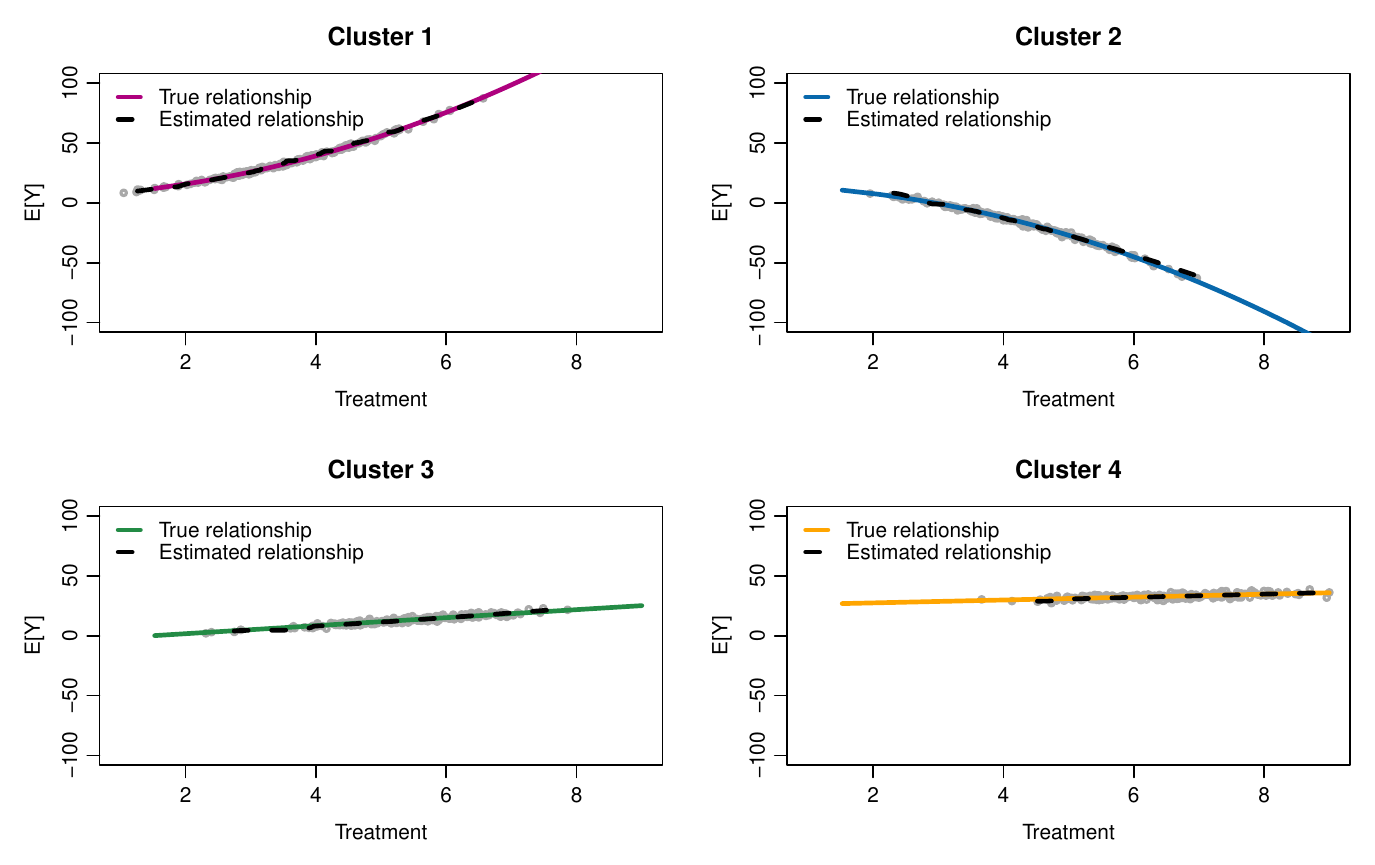} 
    \caption{DRFs estimated using the Cl-DRF approach compared to the true simulated relationships. The colored solid lines represent the four simulated relationships for all values of treatment as in equation \ref{simulexample}; the dashed black lines are the estimated DRFs by using the Cl-DRF approach.}
    \label{fig:cldrfs}    
\end{figure}

\noindent Let us consider the motivating example discussed in Section \ref{Sec:motivating}. We now apply the Cl-DRF estimator and the results are shown in Figure \ref{fig:cldrfs}. To apply the Cl-DRF estimator we need to choose the number of clusters $C$. Considering different values in the set $C = \{1,2,3,4,5,6,7\}$, we compute the values of the BIC-like criterion as shown in \eqref{eq:bic}. As shown in Figure \ref{fig:bicex}, an elbow is detected at $C=4$, which is the true number of clusters. The Rand Index\footnote{The Rand Index measures the similarity between two data clusterings by assessing the proportion of pairs of elements that are consistently grouped together or separately in both partitions \citep[see][]{rand1971objective}. Once a true partition is available, the Rand Index can be used to assess the similarity between the obtained partition with the true one. The index ranges from 0 to 1, where 0 indicates that the clusterings do not agree on any pair of points and 1 indicates that the two partitions under comparison are identical. The higher the index, the better the partition.}, which provides a measure of agreement between the obtained partition with the proposed Cl-DRF estimator and the true one, is equal to 0.98, which suggests an almost perfect identification of the simulated clusters. As shown in Figure \ref{fig:cldrfs}, the DRFs reproduced with the Cl-DRF estimator (dashed black line) are very close to the true DRFs (colored lined), demonstrating the efficacy of the Cl-DRF estimator in this example. 

\subsection{Simulation results}
\label{sec:sim}

\noindent In line with the motivating example discussed in Section~\ref{Sec:motivating},
we conduct an extensive simulation study to assess the finite-sample performance of the proposed Cl-DRF estimator.
The simulations are designed to evaluate two key aspects of the methodology. First, we assess the ability of the information-criterion \eqref{eq:bic} to correctly select the number of clusters $C$.
Because cluster membership is unobserved, accurate recovery of the underlying
partition is a crucial prerequisite for meaningful estimation of
cluster-specific dose--response functions.
To this end, we compute the Rand Index between the estimated and true
partitions across repeated simulations, and we examine the distribution of the
selected number of clusters obtained via the IC-based Elbow criterion discussed in Section \ref{section:test}. Second, conditional on the selected number of clusters, we evaluate the ability
of the Cl-DRF estimator to recover the true cluster-specific dose--response
functions.

To this aim, we consider two complementary data-generating scenarios. In the first scenario, shown in Appendix \ref{app:norandom}, treatment assignment depends on observed covariates through a
cluster-specific assignment mechanism, inducing confounding that must be
addressed via the GPS. In the second scenario, shown in Appendix \ref{app:random}, treatment is randomly assigned and therefore independent of
covariates. Comparing these settings allows us to isolate the role of the treatment
assignment model and to verify that the Cl-DRF estimator behaves appropriately
both in the presence and in the absence of confounding. 
More precisely, we design a simulation study that retains the key structural ingredients of the motivating example—namely, the presence of unobserved cluster-specific treatment-response relationships and heterogeneity in covariate distributions across clusters—while simplifying some components to better isolate the core features of the method. Specifically, the simulation design preserves the violation of the positivity assumption: although treatment is generated conditionally on covariates within each cluster, the supports of the covariates differ across clusters, making some treatment values uncommon or even absent in certain subpopulations.
While the positivity violation is present in both settings, in the simulation setting the treatment ranges exhibit more overlap, making cluster boundaries less distinct. This design choice increases the difficulty of the estimation problem, providing a more conservative test of the Cl-DRF estimator’s ability to detect heterogeneity under less favorable, but more realistic, conditions.
Another key distinction involves the functional form of the outcome model. In contrast to the motivating example in Section \ref{Sec:motivating}, here we follow the simplified specification introduced in Appendix \ref{motivating_noconf}, where the outcome depends only on the treatment. Covariates therefore induce confounding solely through the assignment mechanism, but do not enter the outcome equation directly.
Moreover, we simplify matters by assuming linear outcome models within each cluster.

Overall, the simulation results show that the proposed IC-based procedure
selects the correct number of clusters in the vast majority of replications,
and that the Cl-DRF estimator recovers the cluster structure with a
high Rand Index.
Moreover, the estimated dose--response functions closely track the true
cluster-specific relationships, supporting the use of the Cl-DRF approach
for policy-relevant analysis in settings with clustered heterogeneity.

\color{black}

\section{Estimating the impact of the EU Cohesion Policy through the Cl-DRF estimator}
\label{Sec: application}

\noindent In this section, we apply the Cl-DRF estimator to assess the causal impact of the European Union’s Cohesion Policy—the world’s largest regional development program in terms of geographic scope, budget, and longevity.

\subsection{Background}
While the bulk of the Cohesion Policy aims at supporting the growth processes of the poorest EU regions, it also finances projects in richer regions. This results in a situation where all regions receive some treatment, but the intensity of treatment is vastly heterogeneous in the amount of funds. For instance, from 1994 to 2006, the North-Holland region received an annual average per capita transfer close to €9, whereas the Região Autónoma dos Açores (PT) received €773, almost 85 times more \citep{cerqua2018we}. The Cohesion Policy assignment process suggests that regions receiving large amounts of funds are fundamentally different from those receiving few funds. Estimating a single ADRF under these circumstances could result in comparisons between areas with significant disparities in pre-treatment economic growth, per capita GDP, investment rates, and other relevant factors.

Previous literature has already analyzed and demonstrated the presence of heterogeneity of the Cohesion Policy effects with respect to the intensity of treatment \citep{becker2012too, cerqua2018we}. These studies show that transfers enable faster economic growth in the recipient regions, but the transfer intensity exceeds the aggregate efficiency maximizing level. At the same time, other studies \citep{becker2013absorptive, rodriguez2015quality} have found that the effect of public transfers on economic growth depends on the absorptive capacity of recipient regions (proxied by human capital endowments and quality of institutions). Considering these findings together, they suggest that one should ideally account for both the heterogeneity of treatment in terms of transfer intensity and the heterogeneous responses to the same amount of funds.\footnote{\cite{rodriguez2015quality} make an attempt in this direction, but rather than estimating a DRF for different subgroups of regions, they use an approach not grounded in the potential outcomes framework. Indeed, they estimate a simple parametric method in which they model regional economic growth as a function of the amount of funds received, the regional quality of institutions and the interaction among these terms.}

\subsection{Data}

\noindent In this analysis we will focus on the relationship between EU funds and economic growth for the period from 1994 to 2015. The main data source is the Annual Regional Database of the European Commission’s Directorate General for Regional and Urban Policy (ARDECO) dataset, which provides comprehensive socio-economic and demographic data for regions across the European Union. 

In particular, we will focus on the NUTS-2 regional level and use the GDP per capita adjusted at Purchasing Power Standard (PPS) compound annual growth rate from 1994 to 2015 as the dependent variable.

We will control for the 1994 values of the following variables: i) per capita GDP in PPS; ii) population density; iii) share of gross value added (GVA) in the primary sector; iv) average yearly hours worked per employee; v) average yearly compensation per employee; vi) workplace employment rate.
Another key variable for our analysis is the yearly
estimate of the amount of funds per capita received over the programming period \citep[see]{cerqua2023will} that will be used as the treatment intensity variable.

\begin{table}[ht]
\centering
\caption{Descriptive statistics}
\begin{tabular}{lrrrr}
  \hline
 & Mean & SD & Min & Max \\ 
  \hline
Yearly GDP per capita growth rate (1994-2015) & 2.54\% & 0.66\% & 0.52\% & 5.94\% \\ 
Yearly EU funds per capita (1994-2015) & €78.39 & €92.70 & €7.89 & €425.28 \\ 
GDP per capita in PPS in 1994 & €17,599 & €6,383 & €8,564 & €78,912 \\ 
Population density in 1994 & 455.12 & 1,098.42 & 3.18 & 8,414.19 \\ 
Share GVA primary sector in 1994 & 0.029 & 0.031 & 0.001 & 0.178 \\ 
Avg hours worked per empl. in 1994 & 1,718.80 & 213.83 & 607.23 & 2,359.61 \\ 
Avg compensation per empl. in 1994 & €20,057.09 & €6,567.75 & €4,758.00 & €39,916.34 \\ 
Workplace employment rate in 1994 & 42.09\% & 9.98\% & 27.32\% & 145.18\% \\ 
 \hline
 \label{tab: des}
\end{tabular}
%\vspace{-30pt}
\end{table}

\subsection{Results}

\begin{figure}[!htb]
    \centering
    \includegraphics[width=0.7\linewidth]{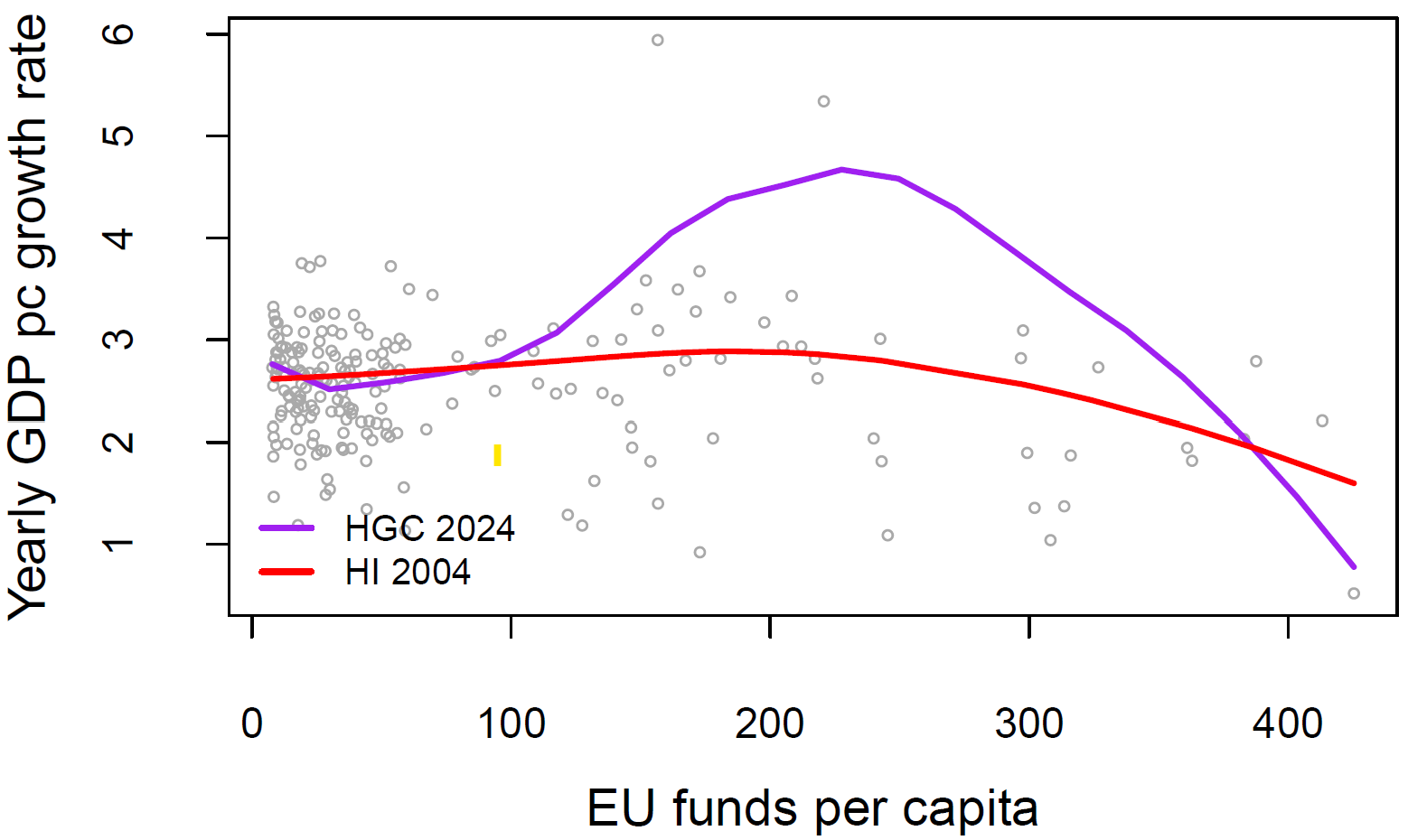}
        \caption{ADRF without cluster structure by using the \citep{hirano2004propensity} approach and following \citep{huling2023independence}. \textit{HI2004} (red line) is the estimated ADRF by using the \citep{hirano2004propensity} approach; \textit{HGC2024} (purple line) is the ADRF estimated by using the \citep{huling2023independence} approach.}
    \label{fig:app_nocluster}    
\end{figure}

\begin{figure}[!htb]
    \centering
    \includegraphics[width=0.7\linewidth]{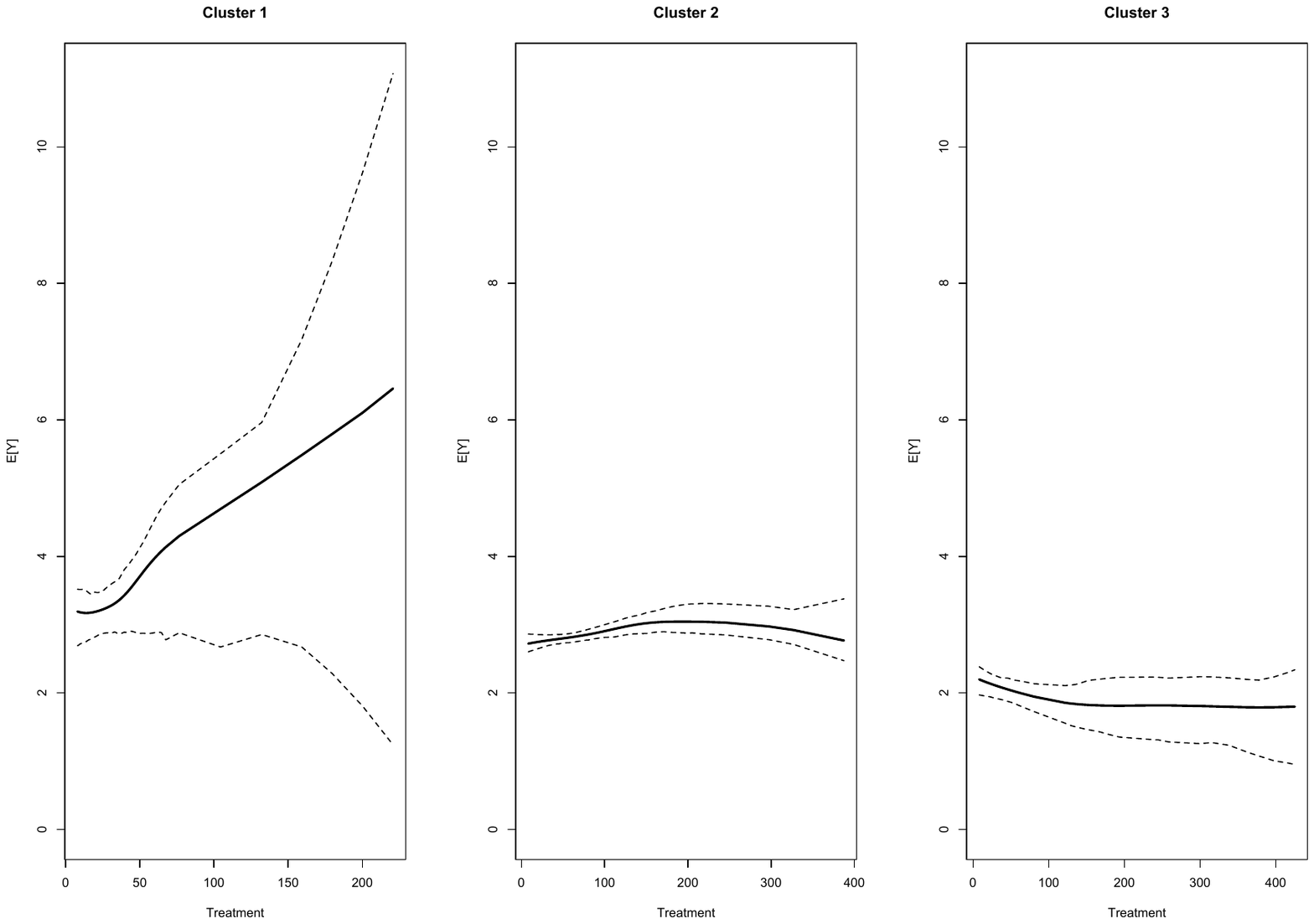}
    \caption{Clustered DRFs with the proposed approach.  95\% confidence intervals are computed using 200 within-cluster bootstrap replications, including GPS estimation at each iteration. The DRF for Cluster 1 is estimated more imprecisely due to the small number of observations in that cluster.}
    \label{fig:app_cldrf}
\end{figure}

\noindent In this section, we estimate the causal relationship between EU funding and economic growth using three approaches for continuous treatments: the GPS of \citet{hirano2004propensity}, the independence weights of \citet{huling2023independence}, and the Cl-DRF estimator. The corresponding DRFs are reported in Figure \ref{fig:app_nocluster} for the first two methods and Figure \ref{fig:app_cldrf} for the Cl-DRF estimator. 

Figure \ref{fig:app_nocluster} shows a slightly positive relationship between EU funds and economic growth, which flattens out for regions receiving more than 100 euros per capita per year and turns even negative for the most subsidized regions. Not surprisingly, this finding is in line with the findings of \citep{becker2012too} who have used the GPS estimator to study the same relationship over the two programming periods 1994-1999 and 2000-2006. 

On the other hand, looking at Figure \ref{fig:app_cldrf} and Table \ref{tab: des clust}, we see that the Cl-DRF algorithm selects three clusters: Cluster 2 exhibits a relationship between treatment intensity and economic growth similar to that reported in Figure \ref{fig:app_nocluster}. In contrast, Cluster 1 shows a steep positive relationship between funds and growth, while regions in Cluster 3 display a slight negative relationship. Notably, the common support of the first cluster is much smaller and does not include any of the most subsidized areas. Among the three clusters, regions in Cluster 1 tend to have higher population densities, higher employment rates, greater wealth, and lower subsidy levels. Importantly, the difference between the DRFs reported in Figure \ref{fig:app_nocluster} and those in Figure \ref{fig:app_cldrf} suggests that the lack of consideration for the likely presence of HTE in the case of continuous treatments might lead to biased estimates and, consequently, to incorrect policy recommendations. For instance, our analysis reveals that, at least for some of the wealthiest EU regions in Cluster 1, there is no evidence supporting the hypothesis that a maximum funding threshold exists beyond which additional resources fail to stimulate further economic growth \citep[see][]{becker2012too, cerqua2018we}. This finding suggests that policymakers can minimize deadweight losses by tailoring subsidy schemes to the heterogeneous marginal returns identified across clusters—avoiding uniform funding caps and aligning allocations with region-specific growth potentials. At the same time, recognizing that some regions cannot fully benefit from additional funding should prompt policymakers to invest in developing those regions’ absorptive capacity.

\begin{figure}[!htb]
    \centering
    \fbox{\includegraphics[width=0.6\linewidth]{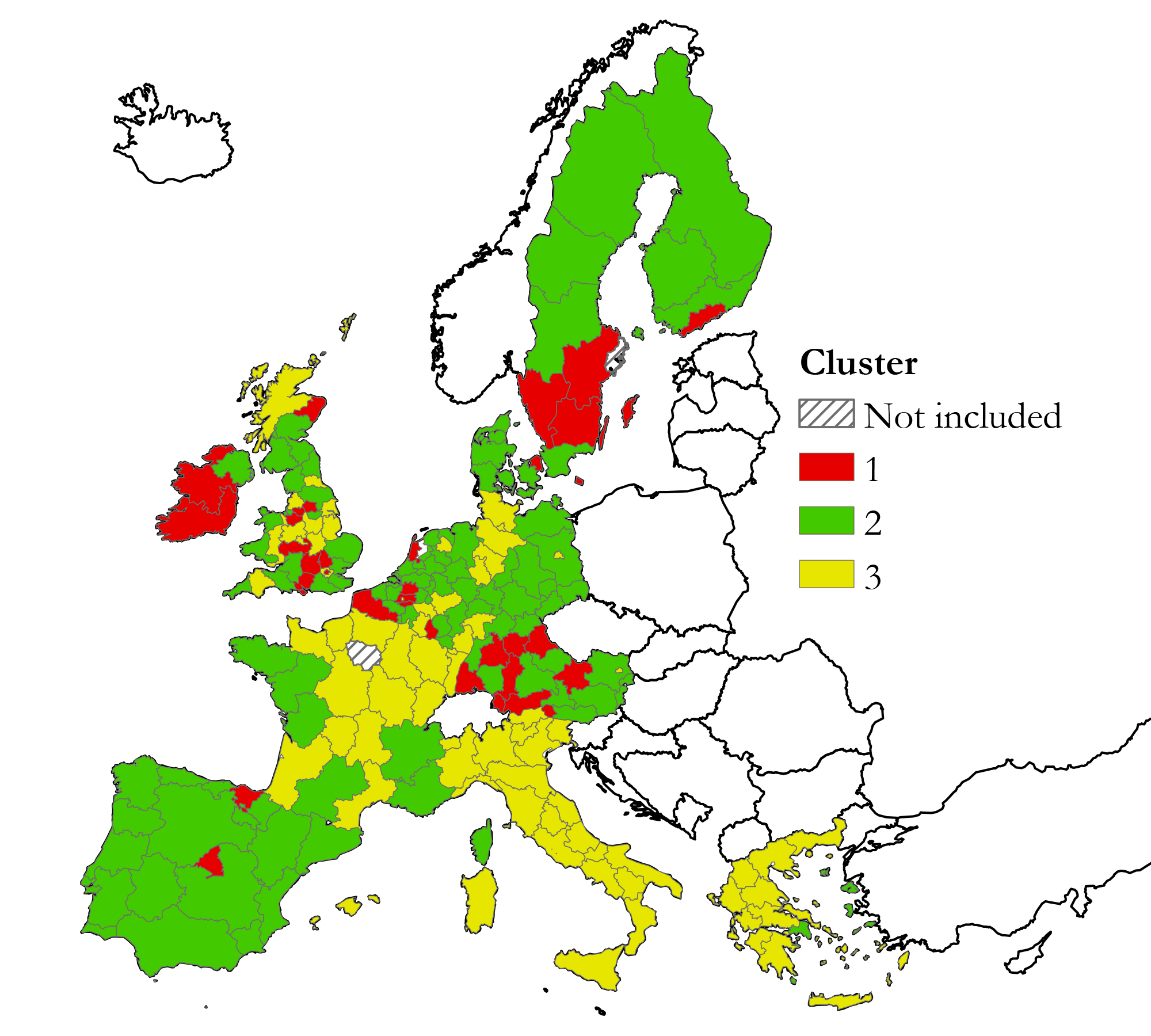}}
    \caption{Map of the three clusters. To limit the issue of extrapolation, we have trimmed from the analysis four regions, the two with the highest/lowest treatment intensity.}
    \label{fig:mappa}
\end{figure}

Looking at the composition of the clusters displayed in Figure \ref{fig:mappa}, we see that Cluster 1 includes regions from Ireland, Southern Germany, and Southern England, while Cluster 2 predominantly consists of regions in Spain, Northern Germany, the Netherlands, Sweden, and Finland. Lastly, Cluster 3 is composed of many French regions as well as nearly all regions in Italy and Greece. These latter countries were among the hardest hit during the Great Recession, which may explain the counterintuitive finding that higher EU funding led to less growth in these regions. Indeed, particularly during the years 2008-2015, regional inequalities in Greece and Italy widened \citep[see][]{fingleton2015shocking}, despite the larger influx of EU funds to their least developed regions.

\begin{table}[ht]
\centering
\caption{Descriptive statistics by Clusters}
\resizebox{0.7\textwidth}{!}{%
\begin{tabular}{lrrr}
  \hline
 & Cluster 1 & Cluster 2 & Cluster 3 \\ 
   \hline
  Yearly GDP per capita growth rate (1994-2015) & 3.24\% & 2.80\% & 1.91\% \\ 
  Yearly EU funds per capita (1994-2015) & €36.13 & €79.59 & €95.16 \\ 
 GDP per capita in PPS in 1994 & €21,689.55 & €16,271.13 & €17,569.48 \\ 
  Population density in 1994 & 824.29 & 271.91 & 535.89 \\ 
  Share GVA primary sector in 1994 & 0.01 & 0.03 & 0.04 \\ 
  Avg hours worked per empl. in 1994 & 1,653.87 & 1,684.22 & 1,792.50 \\ 
  Avg compensation per empl. in 1994 & €23,037.45 & €19,804.13 & €19,095.82 \\ 
  Workplace employment rate in 1994 & 47.79\% & 41.64\% & 40.20\% \\ 
 \\
 \hline
  Number of regions & 33 & 100 & 76 \\ 
   \hline
  \label{tab: des clust}
\end{tabular}
}
\vspace{-30pt}
\end{table}

\section{Conclusions}
\label{Sec: conclusions}

\noindent Continuous treatments are common in applied work, yet existing methods typically target a single population ADRF. This approach is fully appropriate only when treatment responses are homogeneous across units and overlap is sufficient across treatment intensities. In many observational settings, however, treatment effects and treatment support vary systematically across subpopulations. In such cases, pooling units and estimating a single ADRF generally fails to recover the true relationship, potentially leading to misleading policy conclusions. We address these limitations by proposing the Clustered Dose--Response Function  estimator, which replaces population-level identification with cluster-specific DRFs capturing distinct causal response patterns.

By integrating clustering methodologies into the analysis of continuous treatments, this paper expands the methodological toolkit for causal inference in settings characterized by treatment effect heterogeneity and limited overlap. The Cl-DRF estimator provides more detailed and policy-relevant information to decision makers by uncovering distinct causal response patterns that are obscured by population-level analyses. This integrative approach reflects the increasing complexity of policy interventions and responds to the growing demand for more sophisticated tools in policy evaluation, as emphasized in recent contributions to the literature \citep{wu2022matching}. 

The Cl-DRF offers several advantages. First, it relies on weaker versions of the weak unconfoundedness and positivity assumptions by requiring them to hold only within clusters. Second, it enhances interpretability by delivering cluster-specific dose--response relationships. Third, it reduces the scope for specification search and p-hacking by making heterogeneity explicit. Finally, it naturally accommodates data-driven clustering procedures.

We apply the Cl-DRF estimator to evaluate the effects of EU Cohesion Policy in EU-15 NUTS-2 regions. When the clustering structure is ignored, standard estimators suggest a modest positive effect of EU funds on economic growth that plateaus beyond €100 per capita and becomes negative for the most highly funded regions. In contrast, the Cl-DRF uncovers three distinct response patterns: one group exhibits a flat or mildly positive effect, a second displays a strong positive response, and a third shows a slightly negative effect. Ignoring this heterogeneity risks producing biased estimates and misleading policy conclusions. By contrast, our analysis demonstrates that more targeted funding strategies, combined with investments aimed at strengthening regional absorptive capacity, can reduce deadweight losses and enhance policy effectiveness.

This framework opens up several promising directions for future research. In the current implementation, the outcome model is assumed to be parametric and to share the same functional form across clusters. This modeling choice reflects the main focus of the paper, namely the identification of DRFs while allowing for heterogeneity across clusters through cluster-specific estimation of model parameters. While the parameters of the DRF are permitted to vary across clusters, its functional form is specified ex-ante and held fixed. An important avenue for future research is therefore to relax this restriction, either by allowing the functional form itself to differ across clusters or by adopting more flexible, potentially nonparametric, estimation strategies. In addition, the Cl-DRF approach is developed under the working assumption that units can be partitioned into a finite number of clusters characterized by distinct dose--response relationships. While such an assumption is common in the literature on grouped heterogeneity \citep[see, for example,][]{bonhomme2015grouped,sugasawa2021grouped}, developing formal statistical tests to assess its empirical validity represents another important direction for future research.

\section{Data Availability Statement}\label{data-availability-statement}
The authors confirm that the data supporting the findings of this study are available within the article or its supplementary materials.

\bibliography{draft}

\newpage

\begin{appendices}
\section{Proofs}
\label{app1}

\subsection{Lemma 1}\label{app:lemma1}

\begin{proof}
Fix $c\in\{1,\dots,C\}$ and $t\in\mathcal T_c$. By Assumption~B2,
$Y_i(t)\perp T_i\mid X_i, C_i=c$, hence
\[
\mathbb E\!\left[Y_i(t)\mid X_i, C_i=c\right]
=
\mathbb E\!\left[Y_i(t)\mid T_i=t, X_i, C_i=c\right].
\]
By Assumption~A1 (consistency), if $T_i=t$ then $Y_i=Y_i(t)$, and therefore
\[
\mathbb E\!\left[Y_i(t)\mid T_i=t, X_i, C_i=c\right]
=
\mathbb E\!\left[Y_i\mid T_i=t, X_i, C_i=c\right].
\]
Combining the two displays yields
$\mathbb E\!\left[Y_i(t)\mid X_i, C_i=c\right]
=
\mathbb E\!\left[Y_i\mid T_i=t, X_i, C_i=c\right]$.
Taking expectations with respect to the conditional distribution of $X_i$ given
$C_i=c$ and using the law of iterated expectations gives \eqref{eq:gformula_clust}.
\end{proof}

\subsection{Lemma 2}\label{app:coro1}

\begin{proof}[Proof of Lemma~\ref{cor:gps_representation}]
By Lemma~\ref{lem:gformula}, under Assumption A1 and Assumptions~B1--B3,
\[
\mu_c(t)
=
\mathbb{E}\!\left[
\mathbb{E}\!\left(
Y_i \mid T_i=t,\, X_i,\, C_i=c
\right)
\mid C_i=c
\right].
\]
Fix $c$ and consider the conditional distribution given $C_i=c$.
By the balancing property of GPS \citep{hirano2004propensity}, the cluster-specific GPS $R_{i,c}(t)$ is a balancing score for $X_i$ at treatment
level $t$ within cluster $c$.
Moreover, by Assumption~B2 (weak unconfoundedness within clusters),
$Y_i(t)\perp T_i \mid X_i, C_i=c$.
Combining the balancing property of $R_{i,c}(t)$ with weak
unconfoundedness yields
\[
Y_i(t)\perp T_i \mid R_{i,c}(t), C_i=c,
\quad \text{for a.e. } t \in \mathcal T_c.
\]
Therefore,
\[
\mathbb{E}\!\left(
Y_i(t)\mid T_i=t,\, X_i,\, C_i=c
\right)
=
\mathbb{E}\!\left(
Y_i(t)\mid T_i=t,\, R_{i,c}(t),\, C_i=c
\right),
\quad \text{for a.e. } t \in \mathcal T_c.
\]
By Assumption~A1 (consistency), $Y_i=Y_i(t)$ whenever $T_i=t$, and thus
\[
\mathbb{E}\!\left(
Y_i \mid T_i=t,\, X_i,\, C_i=c
\right)
=
\mathbb{E}\!\left(
Y_i \mid T_i=t,\, R_{i,c}(t),\, C_i=c
\right),
\quad \text{for a.e. } t \in \mathcal T_c.
\]
Substituting into the expression from Lemma~\ref{lem:gformula} yields the
claimed representation.
\end{proof}

\subsection{Proposition 1}\label{app:prop1}

\begin{proof}[Proof of Proposition~\ref{prop:updates}]
Fix a cluster $c$ and treat $\{\kappa_{i,c}\}_{i=1}^n$ as given.
Define the $n\times 1$ outcome vector $\mathbf{y}:=(y_1,\ldots,y_n)'$ and the $n\times p$ regressor matrix
$\mathbf{Z} := (\mathbf{z}_{1},\ldots,\mathbf{z}_{n})'$, with $p=3$ in the case of linear specification. If higher order polynomials and/or interactions are included we gey $p>3$.
Let $\boldsymbol{\kappa}_c:=\mathrm{diag}(\kappa_{1,c},\ldots,\kappa_{n,c})$ be the diagonal membership matrix.
Then the contribution of cluster $c$ to the objective function is
\[
J_c(\boldsymbol{\alpha}_c)
=
(\mathbf{y}-\mathbf{Z}\boldsymbol{\alpha}_c)'\boldsymbol{\kappa}_c(\mathbf{y}-\mathbf{Z}\boldsymbol{\alpha}_c).
\]
If $\mathbf{Z}'\boldsymbol{\kappa}_c\mathbf{Z}$ is nonsingular, the first-order condition yields
\[
-2\mathbf{Z}'\boldsymbol{\kappa}_c(\mathbf{y}-\mathbf{Z}\boldsymbol{\alpha}_c)=\mathbf{0},
\]
hence
\[
\widehat{\boldsymbol{\alpha}}_c
=
(\mathbf{Z}'\boldsymbol{\kappa}_c\mathbf{Z})^{-1}\mathbf{Z}'\boldsymbol{\kappa}_c\mathbf{y}.
\]

\noindent Then, fix $\{\boldsymbol{\alpha}_c\}_{c=1}^C$ and the current regressors $\{\mathbf{z}_{i,c}\}$ (hence including the current GPS values).
Because $\sum_{c=1}^C\kappa_{i,c}=1$ and $\kappa_{i,c}\in\{0,1\}$, minimisation of the objective function with respect to
$\{\kappa_{i,c}\}$ decouples across $i$.
For each unit $i$, the term
$\sum_{c=1}^C \kappa_{i,c}\,\bigl(y_i-\boldsymbol{\alpha}_c'\mathbf{z}_{i,c}\bigr)^2
$ is minimised by assigning $i$ to the cluster achieving the smallest squared residual.
Equivalently,
\[
\widehat{\kappa}_{i,c}
=
\mathbf{1}\!\left[
\bigl(y_i-\boldsymbol{\alpha}_c'\mathbf{z}_{i,c}\bigr)^2
\le
\bigl(y_i-\boldsymbol{\alpha}_{c'}'\mathbf{z}_{i,c'}\bigr)^2,\ \forall c'\neq c
\right],
\]
which is \eqref{eq:kappa}.

Finally, given $\kappa$, within each cluster $c$ the treatment assignment model is
\[
T_i \mid X_i,\ C_i=c \sim \mathcal{N}(\boldsymbol{\beta}_c'X_i,\ \sigma_c^2).
\]
Let $\mathcal{I}_c:=\{i:\kappa_{i,c}=1\}$, and denote by $\mathbf{t}_c$ and $\mathbf{X}_c$ the treatment vector and
covariate matrix restricted to $\mathcal{I}_c$.
Under Gaussianity, the maximum-likelihood estimators satisfy
\[
\widehat{\boldsymbol{\beta}}_c
=
(\mathbf{X}_c'\mathbf{X}_c)^{-1}\mathbf{X}_c'\mathbf{t}_c,
\qquad
\widehat{\sigma}_c^2
=
\frac{1}{|\mathcal{I}_c|}
\sum_{i\in\mathcal{I}_c}
\bigl(T_i-\widehat{\boldsymbol{\beta}}_c'X_i\bigr)^2.
\]
The cluster-specific GPS used in the outcome regression is then
the conditional density evaluated at the realised treatment,
\[
\widehat{R}_{i,c}
=
\frac{1}{\sqrt{2\pi\widehat{\sigma}_c^2}}
\exp\!\left(
-\frac{(T_i-\widehat{\boldsymbol{\beta}}_c'X_i)^2}{2\widehat{\sigma}_c^2}
\right),
\qquad i\in\mathcal{I}_c.
\]
Finally, the scalar GPS covariate used for unit $i$ is
\[
\widehat{r}_i
=
\sum_{c=1}^C \kappa_{i,c}\,\widehat{R}_{i,c}.
\]
Combining Steps~1--3 yields the closed-form block updates employed by the Cl-DRF algorithm.
\end{proof}

\section{Motivating example}

\subsection{DRFs estimated using the \texorpdfstring{\citep{huling2023independence}}{Huling et al. (2023)} (HCG) approach within each cluster}
\label{drf_huling}

\begin{figure}[!htb]
    \centering
\includegraphics[width=0.8\linewidth]{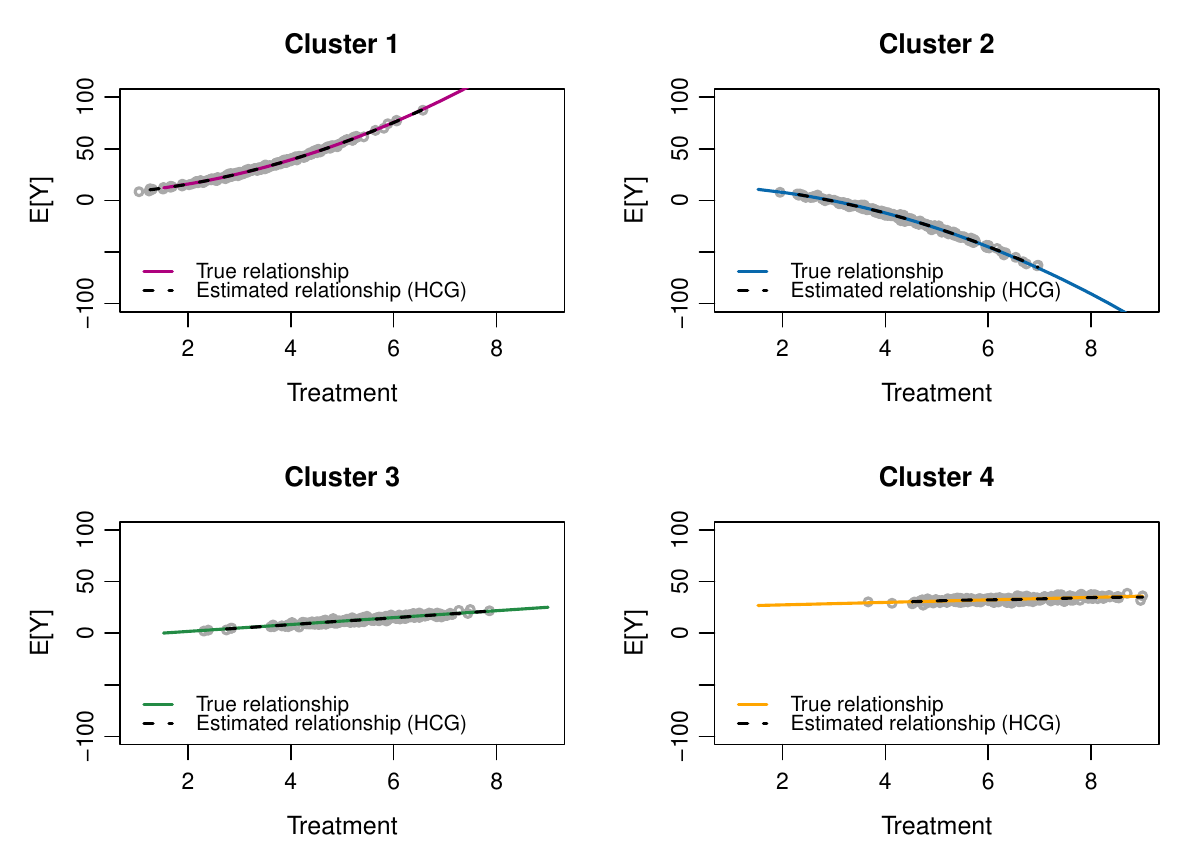} 
\caption{DRFs estimated using the \citep{huling2023independence} approach within each cluster, under the assumption that the cluster structure is known, compared to the true simulated relationships. The colored solid lines represent the four simulated relationships for all values of treatment as in equation \ref{simulexample}; the dashed black lines are the estimated DRFs by using the \citep{huling2023independence} approach.}
    \label{fig:adrfshuling}    
\end{figure}

\subsection{Motivating example without confounding in the outcome model}
\label{motivating_noconf}
In the main text (Section \ref{Sec:motivating}), the outcome is specified as a function of both the treatment and the covariates, allowing covariates to influence the causal effect of the treatment through a treatment–covariate interaction term. This design introduces confounding through two channels: covariates affect the assignment mechanism, and they also directly moderate the treatment effect.

\begin{figure}[!htb]
    \centering
    \includegraphics[width=0.5\linewidth]{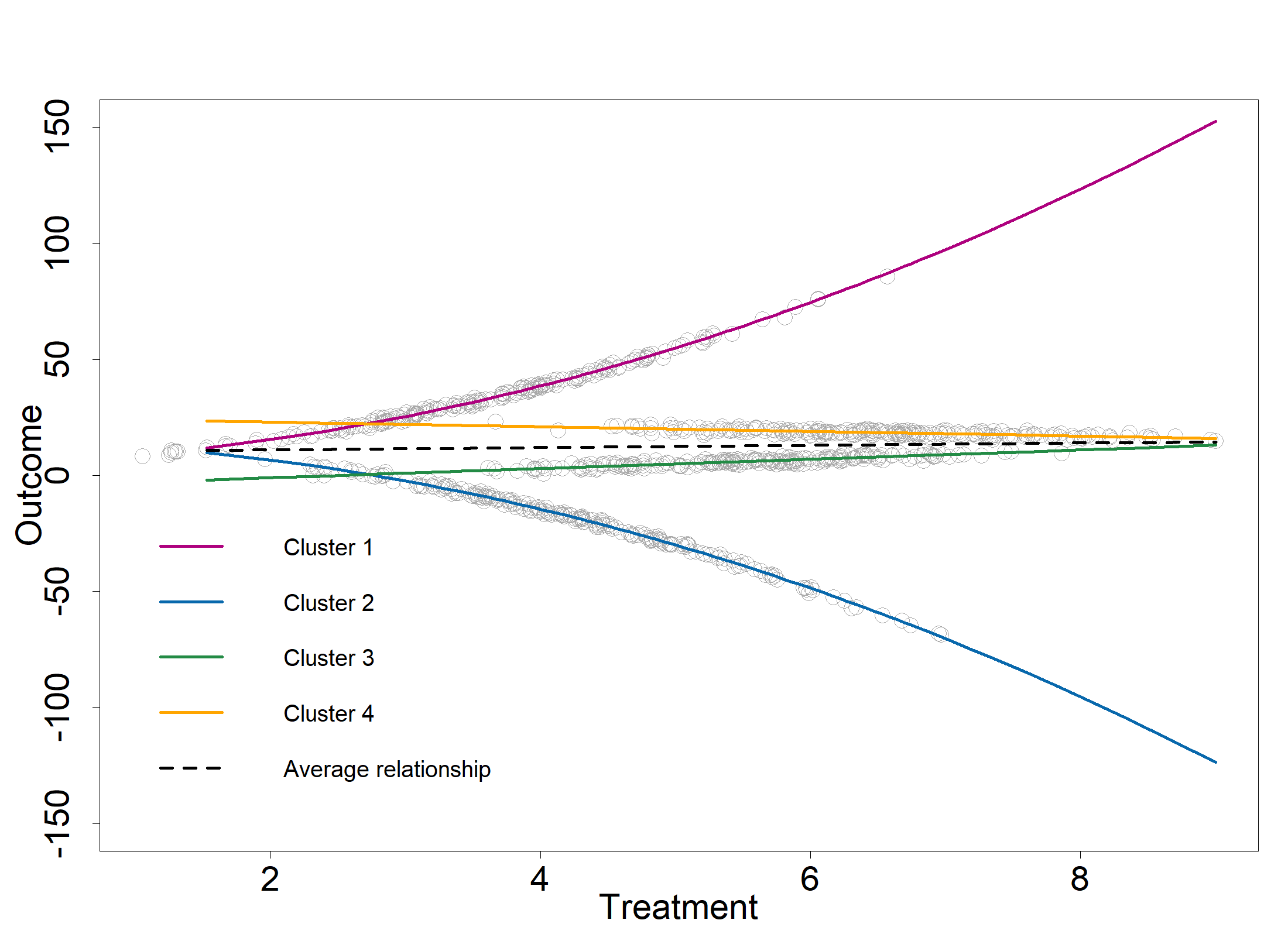}\quad
    \includegraphics[width=0.45\linewidth]{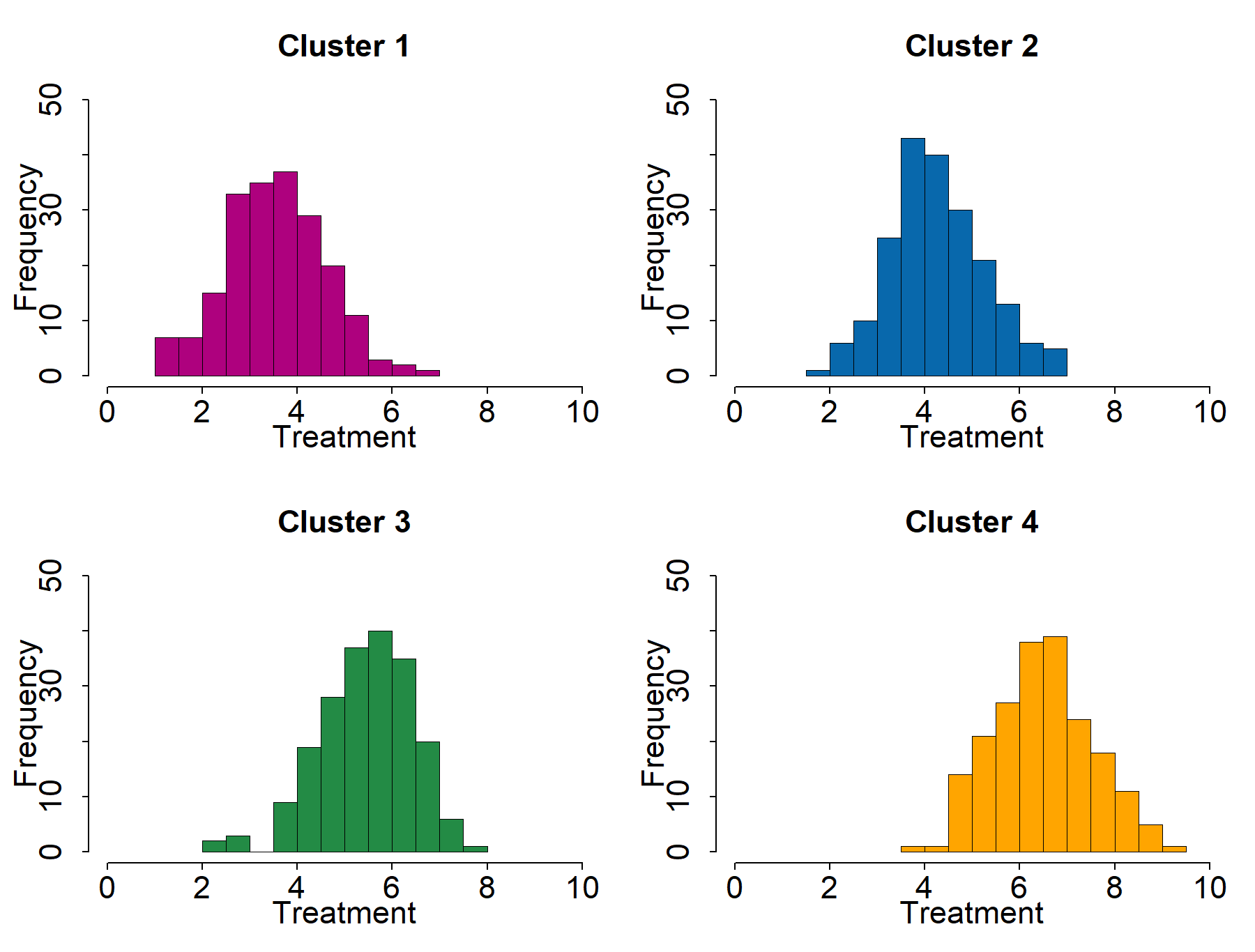} 
    \caption{Relationship between treatment and outcome for all units and at the cluster level (left-hand side) and histogram of distribution of treatment within clusters (right-hand side). Simulated data with $n = 800$ units divided into $C = 4$ equally sized clusters are used. Treatment is assigned based on cluster-specific functions of two pre-treatment covariates, and outcomes are generated using distinct treatment–outcome relationships within each cluster.}
     \label{fig:rel_conf}
\end{figure}

\begin{figure}[!htb]
    \centering  
        \includegraphics[width=0.7\linewidth]{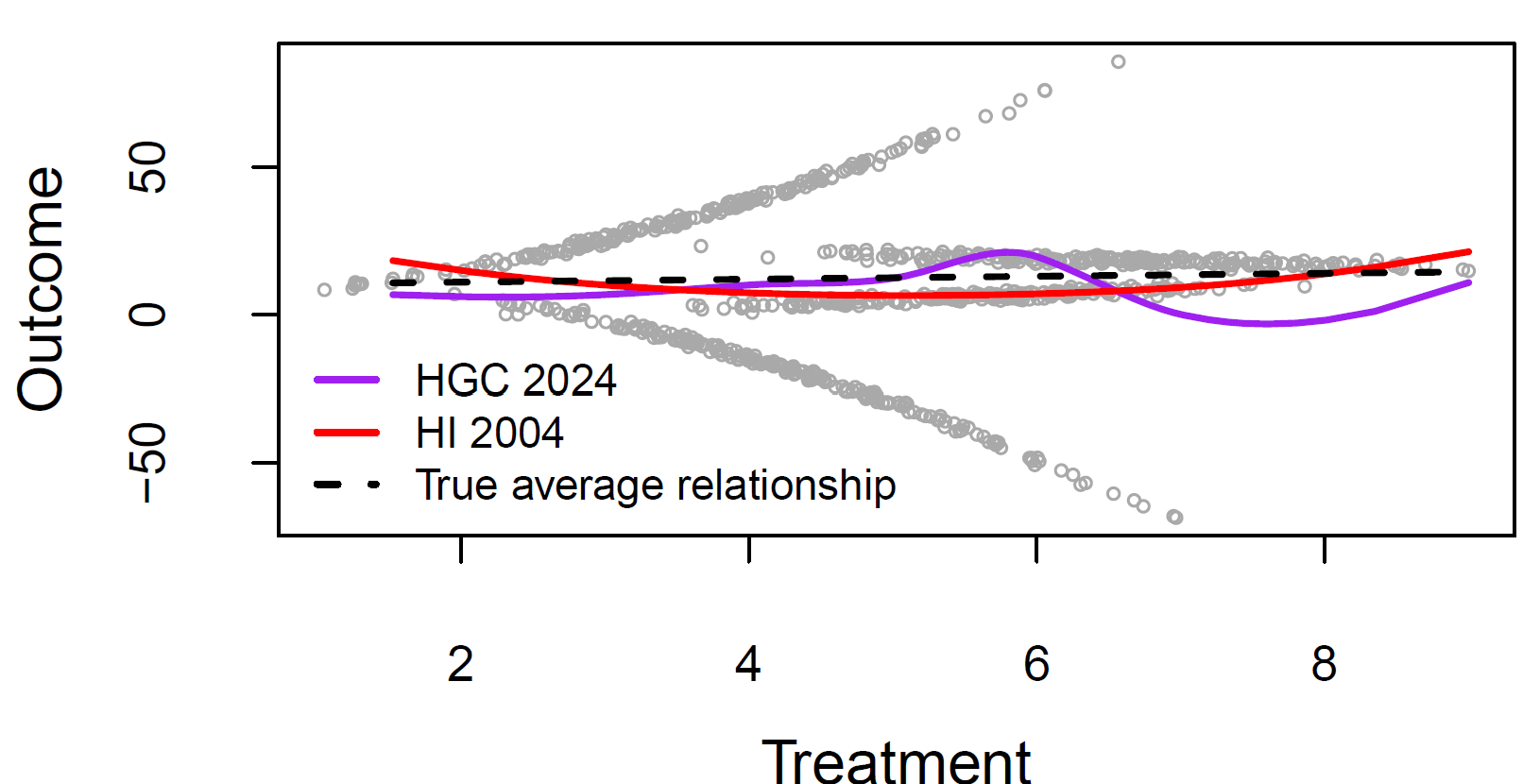}
    \caption{ADRF without cluster structure by using the \cite{hirano2004propensity} approach and following \cite{huling2023independence}. \textit{True average relationship} (dashed black line) represents the average of the four simulated relationships as in equation \ref{simulexample}; \textit{HI2004} (red line) is the estimated ADRF by using the \citep{hirano2004propensity} approach; \textit{HGC 2024} (purple line) is the ADRF estimated by using the \citep{huling2023independence} approach.}
    \label{fig:overall_conf}    
\end{figure}

\begin{figure}[!htb]
    \centering
\includegraphics[width=0.8\linewidth]{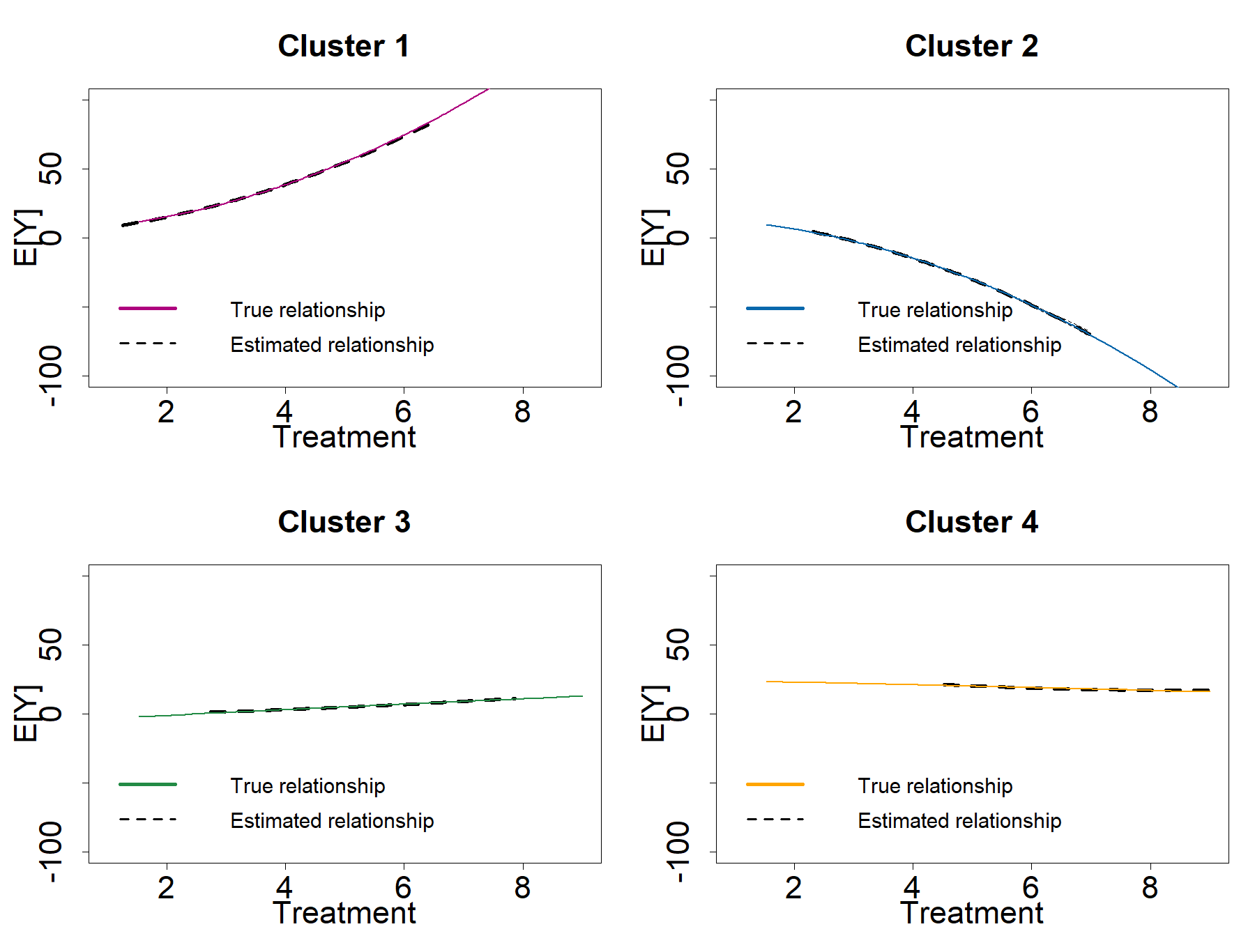} 
\caption{DRFs estimated using the \citep{hirano2004propensity} approach within each cluster, under the assumption that the cluster structure is known, compared to the true simulated relationships. The colored solid lines represent the four simulated relationships for all values of treatment as in equation \ref{simulexample}; the dashed black lines are the estimated DRFs by using the \citep{hirano2004propensity} approach.}
    \label{fig:adrfshirano_onf}    
\end{figure}

In this Appendix, we present an alternative specification in which the outcome is generated as a function of the treatment only. Even though covariates do not enter the outcome equation explicitly, they still induce confounding indirectly through the cluster-specific assignment mechanism, because the treatment depends on covariates. Thus, the outcome remains associated with covariates through the treatment, but no additional direct covariate–outcome pathway is introduced. This simplified specification is useful for illustrating that heterogeneity in the treatment–outcome relationship alone is sufficient to generate the patterns discussed in Section \ref{Sec:motivating}. That is, we assume the following within-cluster relationships
\begin{equation}
\label{simulexample_noconf}
y_i=\begin{cases}
\centering
5+ 2 t_i+ 1.6 t_i^2 + e_{i} & \text { for } c=1\\
15-t_{i}-1.6 t_i^2+e_{i} & \text { for } c=2\\
-5 +2t_i + e_{i} & \text { for } c=3\\
25-1t_{i}+e_{i} & \text { for } c=4\\
\end{cases}
\end{equation}
where $e_i \sim N(0,1)$, $\forall i$. Figure \ref{fig:rel_conf} illustrates the association between treatment and outcome for each cluster. The dashed black line represents the averaged relationship between outcome and treatment for the average unit (without considering the clustering information), i.e., the ADRF.
Then, the estimated cluster-based DRFs are displayed in Figure \ref{fig:overall_conf}. 
Let us assume knowledge of the partition of the units into four groups as in Figure \eqref{fig:rel_conf}, that is, we know exactly what units share similar treatment-outcome relationships. Under this assumption, we can apply the \cite{hirano2004propensity} approach within each of these known groups to estimate cluster-specific DRFs. As shown in Figure \ref{fig:adrfshirano_onf}, the estimated cluster-specific DRFs (dashed black lines) perfectly resemble the true relationships (colored lines) within each cluster.

\section{Simulation study}
\label{app:simulations}

\subsection{No random treatment}\label{app:norandom}
\noindent In what follows we consider the results of a simulation study to evaluate the properties of the proposed Cl-DRF estimator. In particular, we evaluate the appropriateness of the approach discussed in Section \ref{section:test} for choosing the number of clusters $C$. Moreover, we evaluate the DRF reconstruction properties of the Cl-DRF estimator.

Let us consider $500$ replications consisting of datasets with two different sample sizes, that is $n=400$ and $n=800$ units, that are clustered into $C$ equally sized groups. We consider the results of the simulation with $C=4$ and $C=5$ 
In each of the $C$ clusters, we simulate two covariates $\mathbf{x}_i = [x_{i1}, x_{i2}]$  according to the following relationships
\begin{equation}
\label{eq:part1sim}
    t_{i} \mid \mathbf{x}_i, c \sim N\left(\boldsymbol{\beta}'_c \mathbf{x}_i, 1\right),
\end{equation}
where $\boldsymbol{\beta}'_c \neq \boldsymbol{\beta}'_{c^{\prime}}$, $\forall c^{\prime} \neq c$. Then, we simulate the outcome variable as a function of the simulated treatment and assume the following within-cluster relationships
\begin{equation}
\label{part2sim}
y_i=\alpha_{0,c} + \alpha_{1,c} t_{i} +e_{i},
\end{equation}
where $e_i \sim N(0,1)$, $\forall i$. Moreover, we assume that the treatment outcome relationships are different in the $C$ clusters, that is, $\alpha_{0,c} \neq \alpha_{0,c^{\prime}}$  and $\alpha_{1,c} \neq \alpha_{1,c^{\prime}}$  $\forall c^{\prime} \neq c$. 
We assume $x_{1i} \sim U [0, 0.4]$ and $x_{2i} \sim U [0, 0.5]$ for Cluster 1, $x_{1i} \sim U [0.2, 0.6]$ and $x_{2i} \sim U [0.3, 0.6]$ for Cluster 2, $x_{1i} \sim U [0.5, 0.8]$ and $x_{2i} \sim U [0.5, 0.9]$ for Cluster 3 and $x_{1i} \sim U [0.7, 1]$ and $x_{2i} \sim U [0.7, 1]$ for Cluster 4, and $x_{1i} \sim U [0.9, 1]$ and $x_{2i} \sim U [0.9, 1]$ for Cluster 5, where $U[a,b]$ denotes the continuous uniform distribution with support $[a,b]$. Moreover, for the relationships \eqref{eq:part1sim} we simulate the following cluster-specific parameters $\boldsymbol{\beta}_1 = [1.7,2]$, $\boldsymbol{\beta}_2 = [1.2,1.2]$, $\boldsymbol{\beta}_3 = [0.7,0.5]$ and $\boldsymbol{\beta}_4 = [0.5,0.2]$, and $\boldsymbol{\beta}_5 = [0.4,0.1]$ (if $C=5$). Then, we simulate the outcome variable as a linear function of the simulated treatment \eqref{part2sim} and assume the following within-cluster relationships 
\begin{equation}
\label{sim1g4}
y_i=\begin{cases}
\centering
5+  t_i+ e_{i} & \text { for } c=1\\
15-2t_{i} + e_{i} & \text { for } c=2\\
-5  -0.01 t_i + e_{i} & \text { for } c=3\\
25+2 t_i + e_{i} & \text { for } c=4 \\
-20+10 t_i + e_{i} & \text { for } c=5 \ \text{(only if \( C = 5 \))} \\
\end{cases}
\end{equation}
where $e_i \sim N(0,1)$, $\forall i$.

\begin{figure}[!htb]
    \centering
     \begin{subfigure}[b]{0.49\linewidth}
        \centering
        \includegraphics[width=\linewidth]{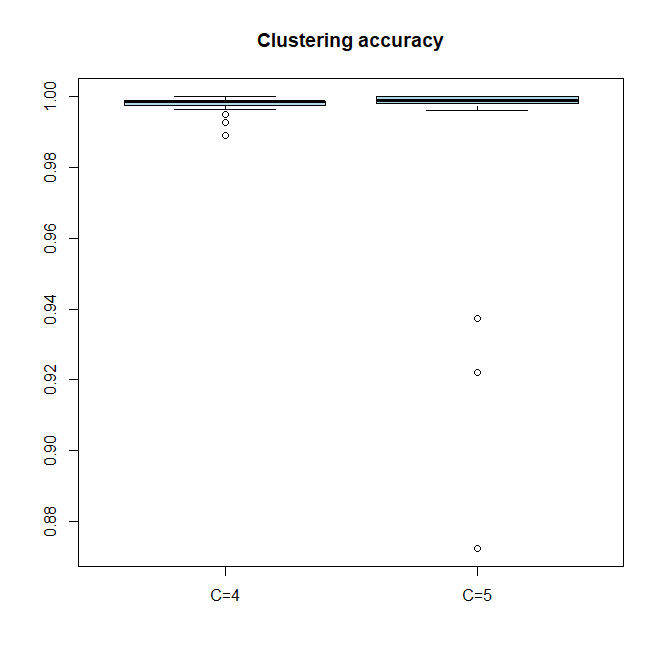}
        \caption{Rand Index for the partitions obtained using the Cl-DRF estimator.}
        \label{fig:simg4ri}
    \end{subfigure}
    \hfill
    \begin{subfigure}[b]{0.49\linewidth}
        \centering
        \includegraphics[width=\linewidth]{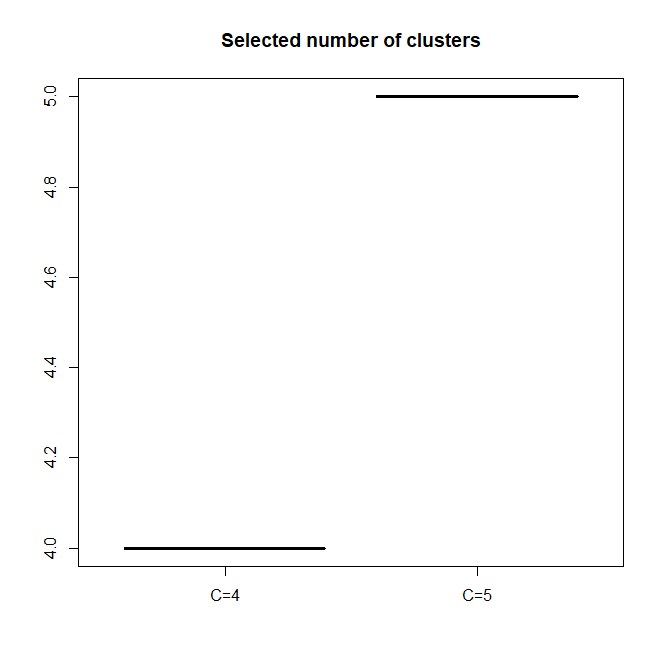}
        \caption{Selected number of clusters.}
        \label{fig:simg4gsel}
    \end{subfigure}
    \caption{Boxplot showing the clustering accuracy, computed with the Rand Index, over the $500$ simulations. The Elbow criterion with the IC criterion \eqref{eq:bic} is used for selecting $C$. Simulated scenario with $C=4$ and $C=5$.}
    \label{fig:combinedsim1}
\end{figure}

\begin{figure}[!htb]
    \centering
   \includegraphics[width=\linewidth]{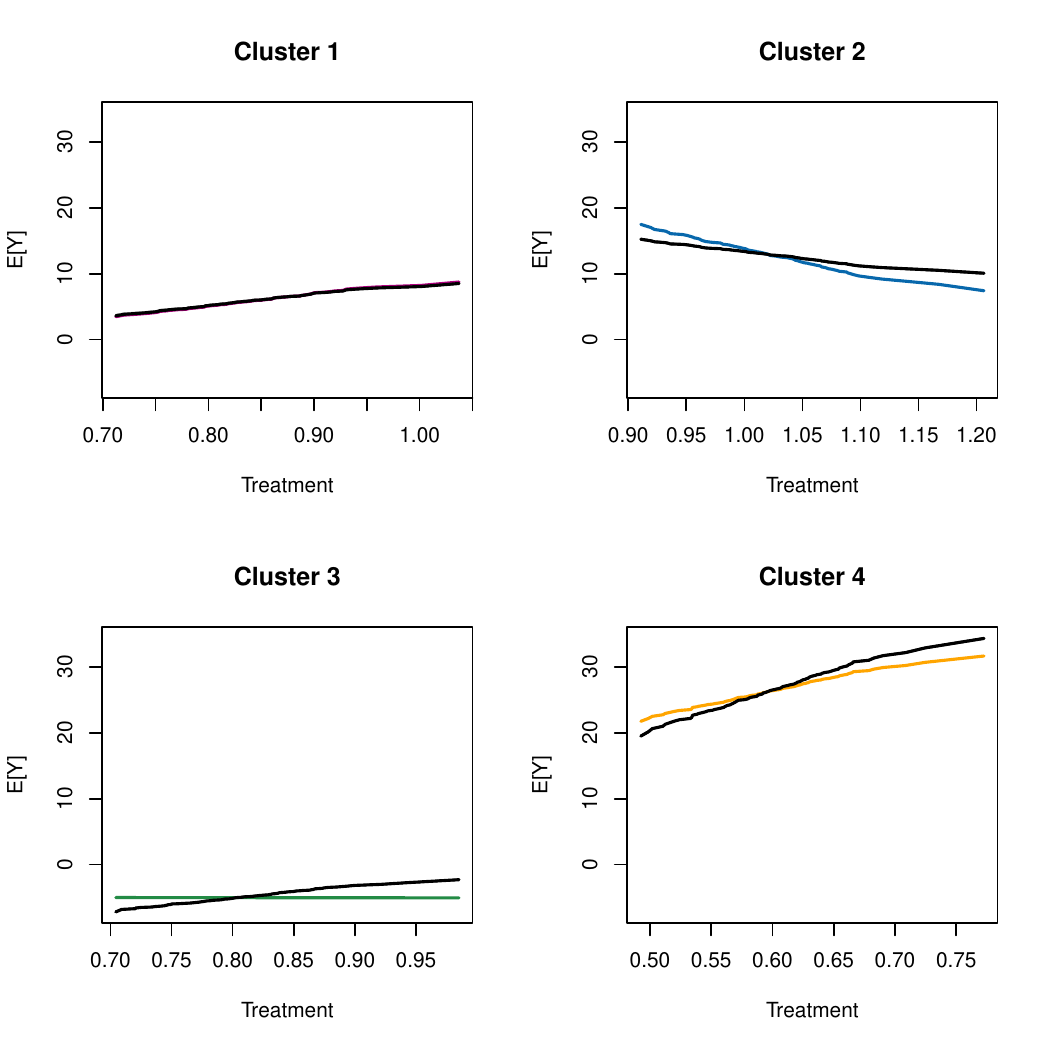}
        \caption{DRFs estimated using the Cl-DRF approach compared to the true simulated relationships. Simulated Scenario with $C=4$ and $n=800$. The colored lines represent the four simulated relationships for all values of treatment as in equation \ref{sim1g4}; the black lines are the estimated DRFs by using the Cl-DRF approach.}
    \label{fig:g4drf800}
\end{figure}

\begin{figure}[!htb]
    \centering
    \includegraphics[width=\linewidth]
    {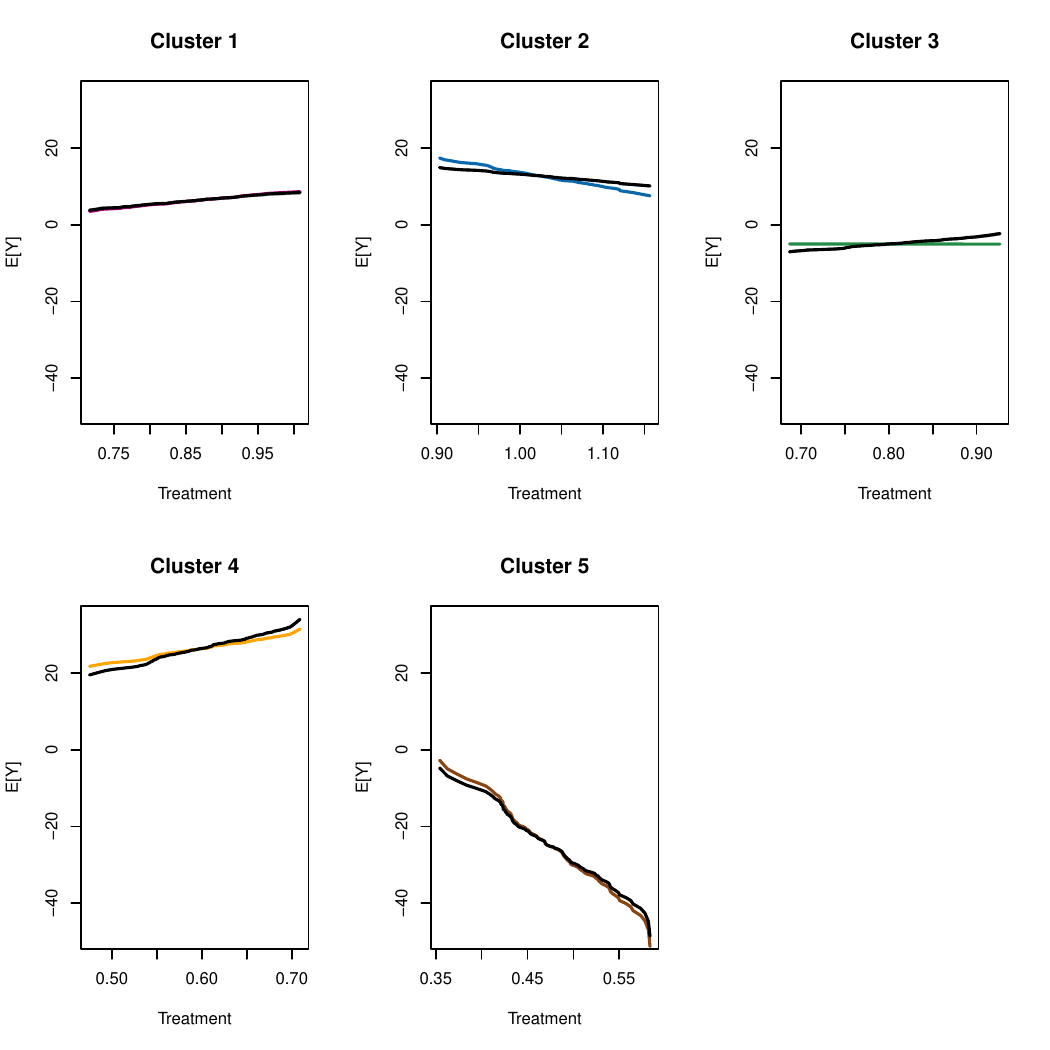}
    \caption{DRFs estimated using the Cl-DRF approach compared to the true simulated relationships. Simulated Scenario with $C=5$ and $n=800$. The colored lines represent the five simulated relationships for all values of treatment as in equation \ref{sim1g4}; the black lines are the estimated DRFs by using the Cl-DRF approach.}
    \label{fig:g5drf800}
\end{figure}

\begin{figure}[!htb]
    \centering
    \includegraphics[width=\linewidth]
    {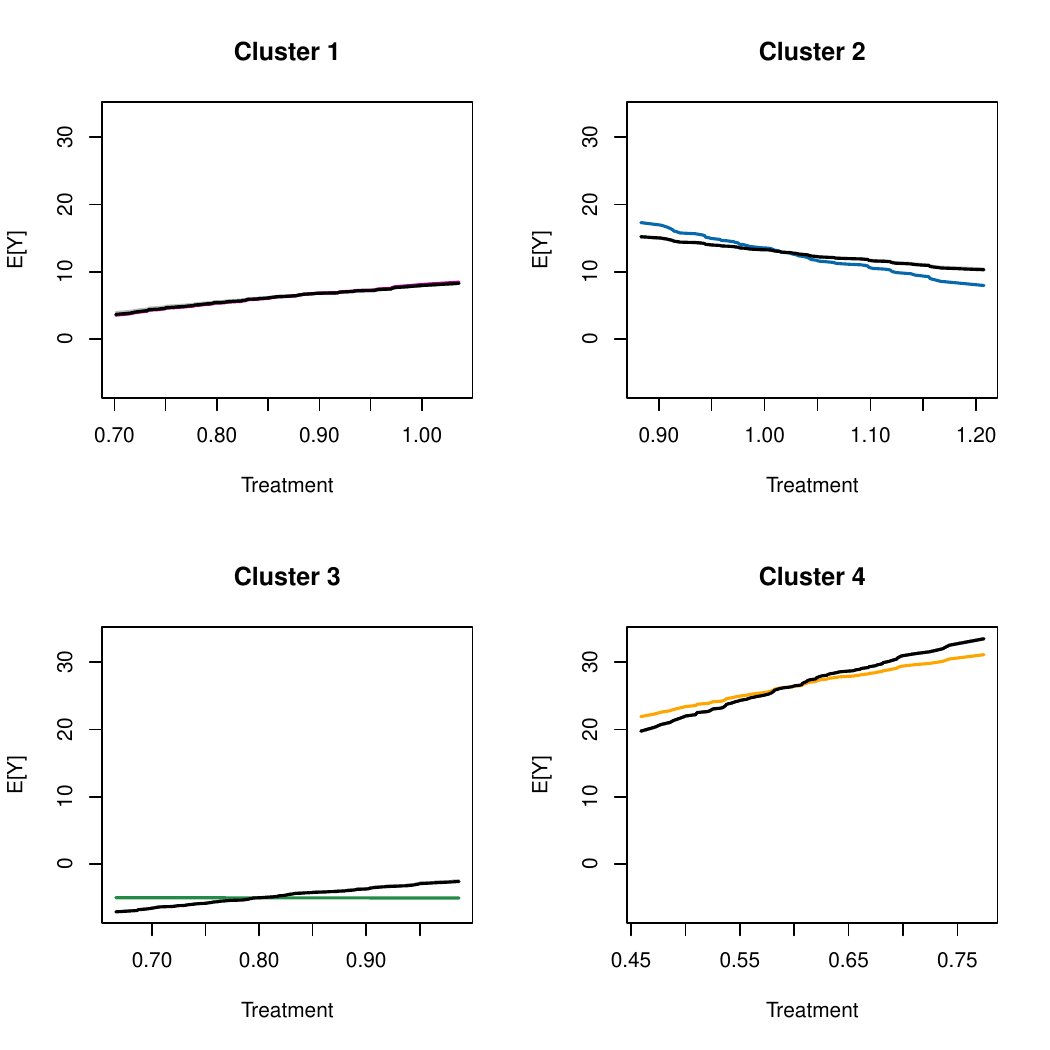}
         \caption{DRFs estimated using the Cl-DRF approach compared to the true simulated relationships. Simulated scenario with $C=4$ and $n=400$. The colored lines represent the four simulated relationships for all values of treatment as in equation \ref{sim1g4}; the black lines are the estimated DRFs by using the Cl-DRF approach.}
    \label{fig:g4drf400}
\end{figure}

\begin{figure}[!htb]
    \centering
    \includegraphics[width=\linewidth]
    {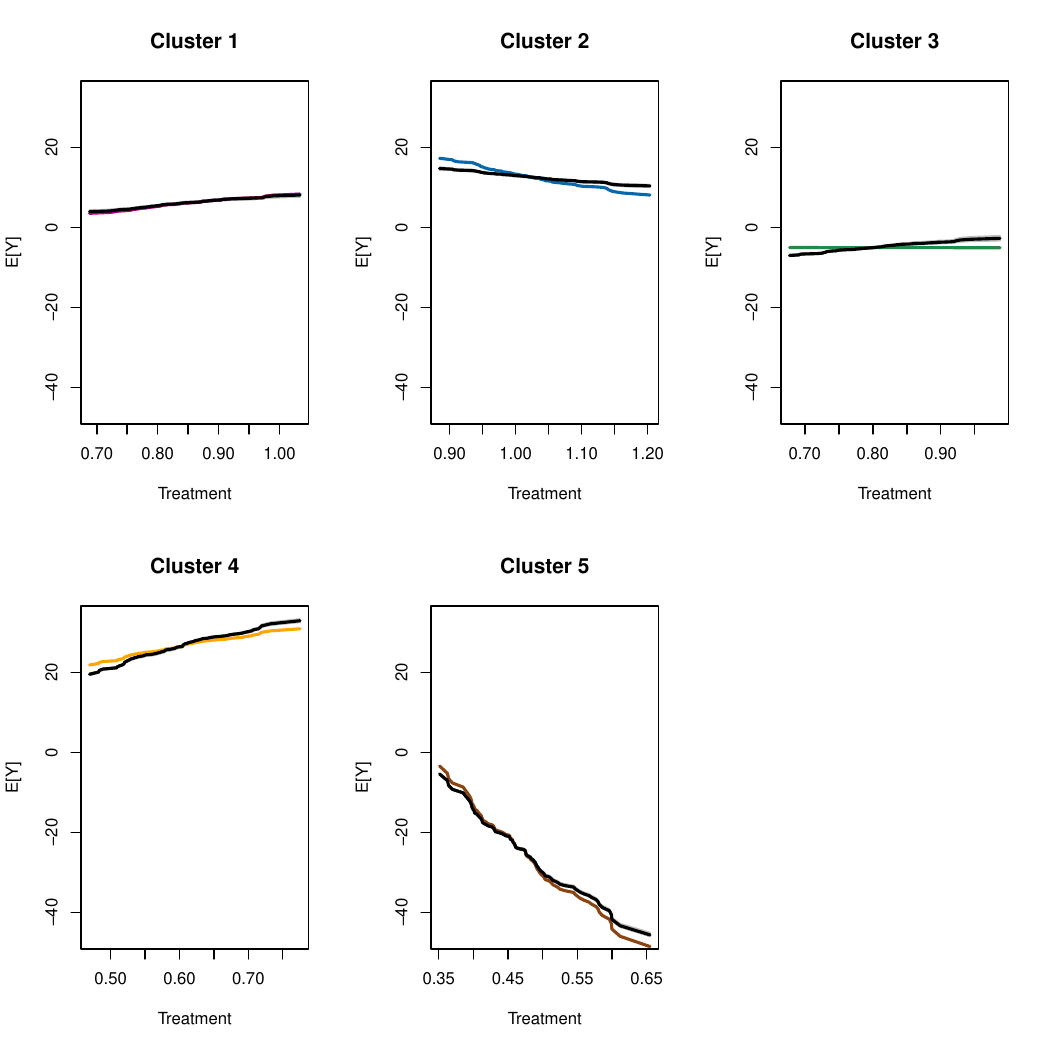}
    \caption{DRFs estimated using the Cl-DRF approach compared to the true simulated relationships. Simulated Scenario with $C=5$ and $n=400$. The colored lines represent the five simulated relationships for all values of treatment as in equation \ref{sim1g4}; the black lines are the estimated DRFs by using the Cl-DRF approach.}
    \label{fig:g5drf400}
\end{figure}

Before applying the Cl-DRF estimator, we need to choose the number of clusters $C$. Following Section \ref{section:test}, we compute the IC criterion \eqref{eq:bic} for different values of $C$ and choose the optimal number of clusters using the data-driven approach discussed in \eqref{eq:bic}, which is based on the Elbow criterion. Figure \ref{fig:combinedsim1} shows the boxplots of the clustering accuracy, computed with the Rand Index, and of the selected number of clusters $C$ across $500$ simulations. 
The results in the right-hand of Figure \ref{fig:combinedsim1} shows that, across all $500$ simulations, the procedure selects the correct number of clusters every time. As a result, the average Rand Index is close to 1 (see the left-hand of the figure), indicating that the Cl-DRF estimator exhibits near-perfect cluster-reconstruction ability. In other words, Figure \ref{fig:combinedsim1} suggests that the approach adopted for the number of clusters selection is appropriate and that, if the correct number of clusters is specified, the Cl-DRF reconstructs the ground truth. This result holds for both scenarios with $n=800$ and $n=400$, although the variability of the results in the setting with $n=800$ is lower compared with $n=400$ and the average Rand Index is slightly higher. This pattern is consistent across both $C = 4$ and $C = 5$. Next, we evaluate the property of the Cl-DRF approach in estimating relevant clustered DRFs from the policy perspective. The DRF for each of the $C=4$ clusters and $C=5$ clusters, in the case of $n=800$ is shown in Figure \ref{fig:g4drf800} and \ref{fig:g5drf800}, while the case of $n=400$ is shown in Figures \ref{fig:g4drf400} and \ref{fig:g5drf400}.
In sum, the Cl-DRF estimator results in a perfect overlap of the DRF for Cluster 1 and for Cluster 5 (if $C=5$), whereas the DRFs for the other clusters are not reconstructed perfectly. However, from the policy perspective, we find that the DRFs estimated with the Cl-DRF estimator are relevant and provide a reliable picture of the differences within each cluster in terms of dose-response relationship. Indeed, we correctly reconstruct that the treatment has a positive relationship with the outcome for the units belonging to Cluster 1 and Cluster 4, while the relationship is negative for those placed in Cluster 2 and Cluster 5. We also correctly find that the treatment-outcome relationship is near zero for units in Cluster 3. 

\subsection{Random treatment}
\label{app:random}

\noindent In what follows we show the results of a simulation study considering random treatment. In particular, we assume that the treatment is not given by any covariate, but $t_i \sim N(1,1)$ $\forall i=1,\dots,n$. Then, we simulate the outcome variable as a function of the simulated treatment according to the following within-cluster relationships
\begin{equation}
\label{sim2app}
y_i=\begin{cases}
\centering
5+  t_i+ e_{i} & \text { for } c=1\\
15-2t_{i} + e_{i} & \text { for } c=2\\
-5  -0.01 t_i + e_{i} & \text { for } c=3\\
25+2 t_i + e_{i} & \text { for } c=4\\
-20- 10 t_i + e_{i} & \text { for } c=5\\
\end{cases}
\end{equation}
where $e_i \sim N(0,1)$, $\forall i$. We again consider two distinct simulation results, where $C=4$ and $C=5$. For the $C=5$ we consider all the relationships in \eqref{sim2app}, while for $C=4$ we consider the first four. We show the results of the comparison between true DRFs and those obtained with the proposed approach, assuming $n=800$, in Figures \ref{fig:sim2g4drf800}--\ref{fig:sim2g5drf800}. 

In sum, the results are in line with those discussed in Appendix \ref{app:norandom}. The approach adopted for selecting the number of clusters leads to very good performances and, under the correct selection of the number of clusters $C$, the Cl-DRF estimator reconstructs the clusters very well. The DRFs estimated by the Cl-DRF estimator are also very close to the true cluster-specific DRFs, suggesting that the method is useful for policy purposes.
\begin{figure}[!htb]
    \centering
    \includegraphics[width=\linewidth]
    {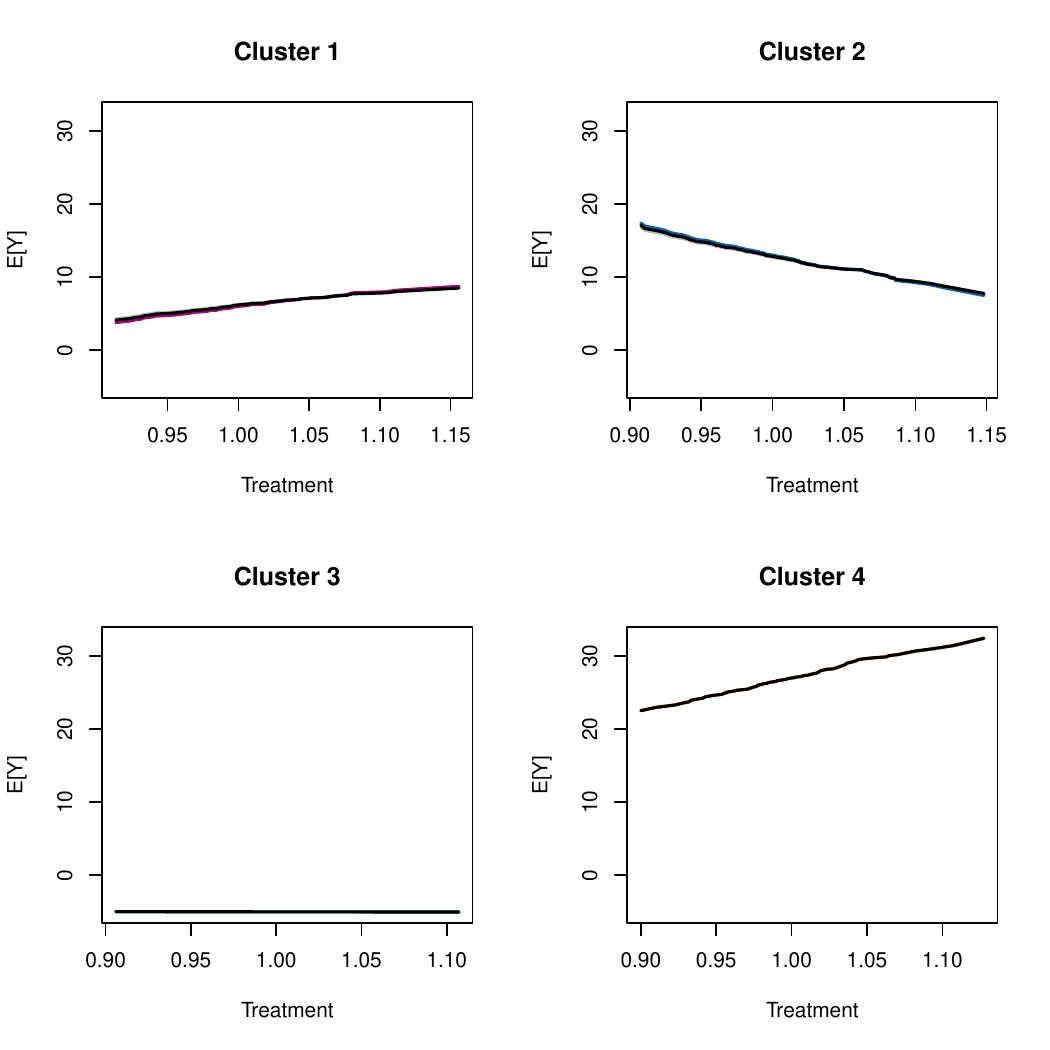}
     \caption{DRFs estimated using the Cl-DRF approach compared to the true simulated relationships.  Simulated Scenario with random treatment. $C=4$ and $n=800$. The colored lines represent the four simulated relationships for all values of treatment as in equation \ref{sim1g4}; the black lines are the estimated DRFs by using the Cl-DRF approach.}
    \label{fig:sim2g4drf800}
\end{figure}

\begin{figure}[!htb]
    \centering
    \includegraphics[width=\linewidth]
    {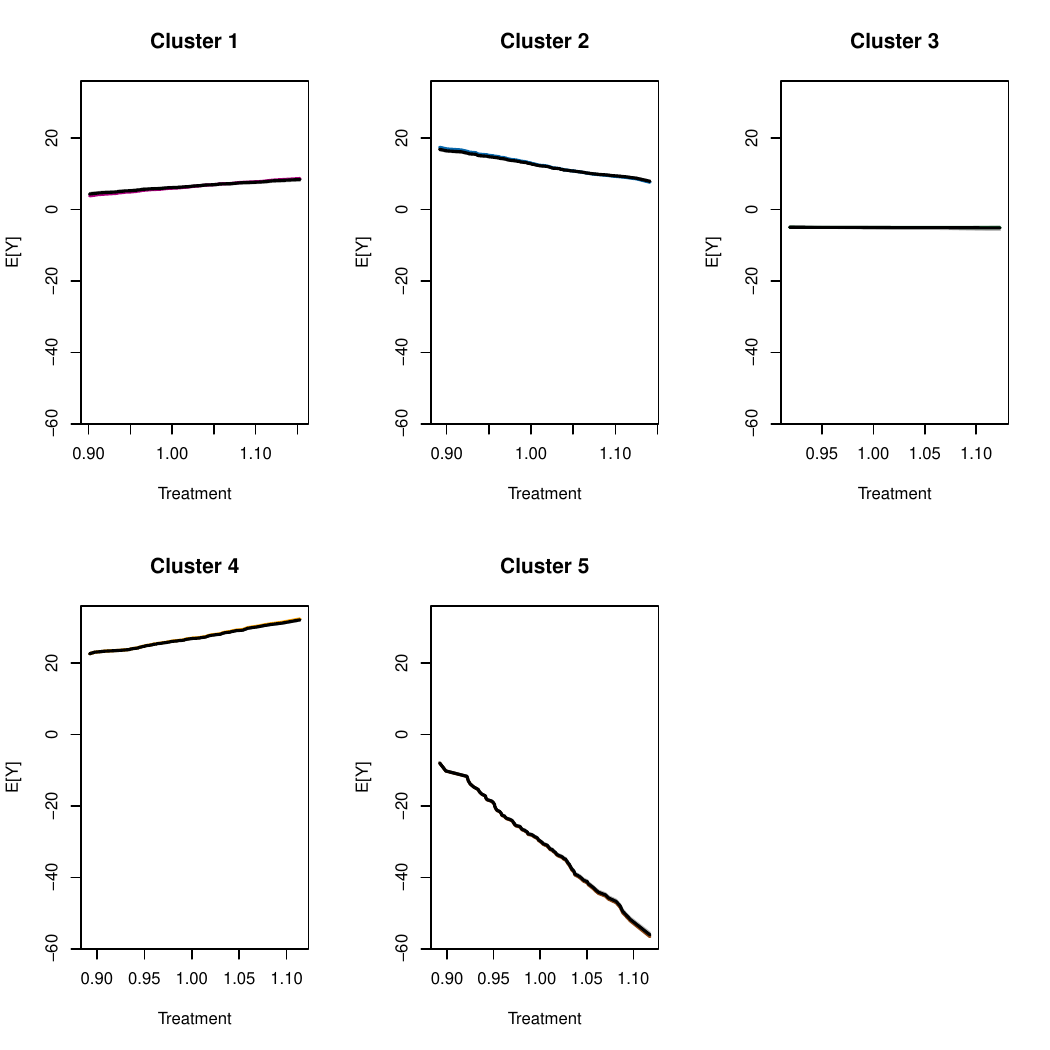}
     \caption{DRFs estimated using the Cl-DRF approach compared to the true simulated relationships. Simulated Scenario with random treatment. $C=5$ and $n=800$. The colored lines represent the four simulated relationships for all values of treatment as in equation \ref{sim1g4}; the black lines are the estimated DRFs by using the Cl-DRF approach.}
    \label{fig:sim2g5drf800}
\end{figure}

\end{appendices}

\end{document}